\documentclass[11pt,a4paper]{article}
\usepackage[T1]{fontenc}
\usepackage[margin=25mm]{geometry}
\usepackage{amsmath,amssymb,mathrsfs,booktabs,graphicx,microtype}
\usepackage{xcolor}
\usepackage{physics}
\usepackage{subfig}
\usepackage[numbers,square]{natbib}
\usepackage[colorlinks=true,linkcolor=blue!55!black,citecolor=blue!55!black,urlcolor=blue!55!black]{hyperref}
\hypersetup{pdftitle={A Path Integral Method with Far-Field Expansion for Open Quantum Dynamics},
 pdfauthor={Zhenning Cai},
 pdfsubject={First- and second-order derivation and numerical results}}
\numberwithin{equation}{section}
\newcommand{\id}{\operatorname{id}}
\newcommand{\dt}{\Delta t}
\newcommand{\Kmax}{K_{\max}}
\newcommand{\Dmax}{D_{\max}}
\newcommand{\Tmax}{T_{\max}}
\newcommand{\ii}{\mathrm{i}}
\title{A Path Integral Method with Far-Field Expansion for Open Quantum Dynamics}
\author{Zhenning Cai\thanks{Department of Mathematics, National University of Singapore,
Level 4, Block S17, 10 Lower Kent Ridge Road, Singapore 119076.
Email: \href{mailto:matcz@nus.edu.sg}{\texttt{matcz@nus.edu.sg}}.}}
\date{November 26, 2026}

\begin{document}
\maketitle
\begin{abstract}
We establish a connection between the quasi-adiabatic propagator path integral (QuAPI) method and fast resummation of Dyson series (FRODS). Building on this connection, we introduce a path integral method with far-field expansion (PIFFE) for quantum systems coupled linearly to a thermal Gaussian bath. PIFFE combines QuAPI with a Taylor expansion of the far-field bath influence, keeping interactions within $r$ time steps in exponential form. The scheme has error $O(\Delta t^2)$ at fixed final time and is free of stochastic or tensor-compression errors. Spin--boson tests show that PIFFE gives results much closer to QuAPI than FRODS does, even for small $r$, while requiring substantially less storage than QuAPI. This allows much longer bath memory lengths and simulations of multilevel systems. Truncating the bath memory and limiting the total auxiliary degree further reduce storage and computational costs.
\end{abstract}

\section{Introduction}
\label{sec:intro}
Open quantum systems describe the dynamics of a quantum system interacting
with its surrounding environment. Such interactions lead to dissipation and
decoherence, whose accurate description is important in quantum
information~\cite{shor1995}, quantum optics~\cite{brune1996}, and chemical
physics~\cite{ishizaki2009}. Since the environment usually contains
many degrees of freedom, directly solving the evolution of the combined
system is computationally expensive. A widely used model represents the
environment by a thermal bath of harmonic oscillators coupled linearly to
the system~\cite{caldeira1983}. In this setting, the bath degrees of freedom can be integrated
out analytically through the Feynman--Vernon influence
functional~\cite{feynman1963}. The resulting reduced dynamics nevertheless
depends on the history of the system. Resolving this dependence without
excessive computational and storage costs is a central difficulty in the
numerical simulation of open quantum systems.

Various approaches have been developed to describe these memory effects.
When the bath correlation time is sufficiently short and the coupling is
weak, a Markovian approximation can simplify the
evolution~\cite{davies1974,lindblad1976}. Outside this
regime, numerical methods need to retain the influence of earlier system
states.

The hierarchical equations of motion (HEOM) introduce auxiliary
density operators to represent memory effects and obtain a coupled system
of evolution equations~\cite{tanimura1989,tanimura1990,ishizaki2005,tanimura2020}.
The multiconfiguration time-dependent Hartree (MCTDH) method treats the
system and bath degrees of freedom explicitly, representing the wavefunction as a
sum of products of time-dependent single-particle functions, with both
the expansion coefficients and the functions determined
variationally~\cite{meyer1990}. Its multilayer extension applies this
representation recursively, allowing a more compact description of
correlated dynamics with many bath degrees of freedom~\cite{wang2003}.
The TEDOPA method maps the bath to a one-dimensional chain with
nearest-neighbor interactions using orthogonal polynomials~\cite{chin2010}.
The resulting system--chain dynamics is then propagated using
time-dependent density matrix renormalization group methods, enabling
simulations beyond the weak-coupling regime~\cite{prior2010}.

Stochastic methods provide another way to compute the reduced dynamics.
Diagrammatic Monte Carlo methods expand the evolution operators in a
Dyson series and sample the resulting high-dimensional integrals~\cite{werner2009}.
The oscillatory integrands cause a dynamical sign problem in real-time
calculations. The inchworm algorithm alleviates this problem by
resumming the diagrammatic series and reusing contributions already
computed at earlier times~\cite{cohen2015,chen2017inchworm,cai2020inchworm}.
Another class of stochastic methods uses non-Markovian stochastic
Schr\"odinger equations. In non-Markovian quantum state diffusion, the
system is described by stochastic wavefunctions driven by colored noise,
with memory terms determined by the bath correlation function; the
reduced density matrix is recovered by averaging their outer
products~\cite{diosi1998}. The hierarchy of pure states (HOPS) provides
a practical formulation in terms of coupled stochastic evolution
equations for auxiliary wavefunctions~\cite{suess2014}.

Path integral methods provide a deterministic alternative for harmonic
baths. The quasi-adiabatic propagator path integral (QuAPI) method combines
a splitting of the evolution operator with an explicit evaluation of the
influence functional~\cite{makri1992}. When the bath correlation decays,
one can neglect interactions beyond a finite memory length and propagate
the reduced dynamics iteratively~\cite{makarov1994,makri1995,makri1995numerical}.
However, storing the contributions of
all system paths within this memory window requires an amount of memory
that grows exponentially with the number of retained time steps. This
becomes particularly restrictive for systems with many levels or baths
with slowly decaying correlations. Several methods reduce this cost by
exploiting the structure of the path sum. For example, the time-evolving
matrix product operator (TEMPO) method compresses the history using tensor
network representations~\cite{strathearn2018}. The small matrix path
integral (SMatPI) method reorganizes the path sum into products of small
matrices~\cite{makri2020disentanglement,makri2020,makri2021amplitudes}.
Further developments include an extended-memory
formulation~\cite{makri2021memory} and a tree-based implementation that
computes the kernel matrices through recurrence relations~\cite{wang2024tree}.

Recently, Sun et al. proposed the fast resummation of Dyson series (FRODS) method~\cite{sun2026}, which constructs first- and second-order schemes from a discrete Dyson expansion. Compared to QuAPI, this approach reduces the storage and computational costs for multilevel systems, while the reduction in cost is accompanied by an additional approximation to the system--bath interaction. In FRODS, the interaction exponential is replaced by a truncated Taylor expansion. Although this approximation preserves the prescribed order as the time step tends to zero, its error can be significant at practical time steps when the coupling is strong. Reducing the time step also increases the number of steps needed to resolve a fixed physical memory length, which may offset the savings in storage. This motivates a method that retains the economical  representation of FRODS while treating the stronger influence terms without a Taylor expansion.

In this paper, we develop the path integral method with far-field expansion (PIFFE) to reduce the loss of accuracy caused by the Taylor approximation in FRODS. Building on a connection between QuAPI and FRODS, PIFFE retains QuAPI's exponential treatment of bath interactions between nearby times and applies a Taylor expansion to interactions between more distant times. This aims to improve accuracy while keeping storage and computational costs low. Numerical tests on spin--boson and multilevel systems demonstrate the accuracy and efficiency of the method, including close agreement with QuAPI in the spin--boson comparisons and the ability
to handle longer bath memory lengths and larger systems.

The rest of this paper is organized as follows. Section~\ref{sec:model} introduces the system--bath model and reviews QuAPI and FRODS. Section~\ref{sec:quapi-to-frods} derives the connection between the two methods. The PIFFE construction, its reduced coefficient representation, and its computational complexity are presented in Section~\ref{sec:localresummation}. Section~\ref{sec:truncation} describes techniques to further reduce computational costs by additional approximations. Numerical results are presented in Section~\ref{sec:numerics}, followed by conclusions in Section~\ref{sec:conclusions}.

\section{Preliminaries}
\label{sec:model}
We consider a system, described by a Hamiltonian $H_s$, coupled linearly to a thermal bath of $L$ harmonic oscillators, described by
\begin{displaymath}
  H_b=\frac12\sum_{l=1}^{L}(\hat p_l^2+\omega_l^2\hat q_l^2),
\end{displaymath}
where $\hat p_l$ and $\hat q_l$ are the momentum and position operators for the $l$-th oscillator, and $\omega_l$ is its frequency. The coupling is assumed to be diagonal, so that the total Hamiltonian has the form
\begin{displaymath}
  H=H_s\otimes\id_b+\id_s\otimes H_b+W_s\otimes W_b, \qquad W_b=\sum_{l=1}^{L}c_l\hat q_l,
\end{displaymath}
where $c_l$ is the coupling constant for the $l$-th oscillator. We further assume that the bath is initially in thermal equilibrium at inverse temperature $\beta$, so that the initial density matrix of the combined system is factorized as
\begin{displaymath}
  \rho(0)=\rho_s(0)\otimes \frac{1}{Z} \mathrm{e}^{-\beta H_b}, \qquad Z=\tr_b(\mathrm{e}^{-\beta H_b}),
\end{displaymath}
where $\rho_s(0)$ is the initial density matrix of the system. We are interested in solving the reduced density matrix of the system at time $t$, given by
\begin{equation} \label{eq:reduced_density}
  \rho_s(t)=\tr_b\left(\mathrm{e}^{-\ii H t}\rho(0)\mathrm{e}^{\ii H t}\right).
\end{equation}
For simplicity, below we will assume that $\ket{i}$ is an eigenstate of $W_s$ satisfying $W_s\ket{i}=w_i\ket{i}$, and we consider an $M$-level system, so that the range of $i$ is $1,\ldots,M$.

One classical approach to compute $\rho_s(t)$ is the quasi-adiabatic propagator path integral (QuAPI) method \cite{makri1995}. It is numerically exact and has good convergence properties. The major challenge in QuAPI is the exponential growth of the computational cost with the number of time steps. Recently, a new method called fast resummation of Dyson series (FRODS) has been proposed \cite{sun2026}, which can significantly reduce the storage and computational cost for multilevel systems. In this section, we will provide a brief review of both methods, which serves as a basis for the derivation of the new scheme.

\subsection{Quasi-adiabatic propagator path integral (QuAPI)}
QuAPI uses a symmetric second-order Trotter splitting of the time-evolution operator. Let $\Delta t$ denote the time step and let $N$ be the number of time steps. The resulting approximation is
\begin{equation} \label{eq:Trotter}
  \mathrm{e}^{-\ii H (N \Delta t)} = \left(\mathrm{e}^{-\ii H \Delta t}\right)^N \approx \left[ \left(\mathrm{e}^{-\ii H_s \Delta t/2} \otimes \id_b\right) \left(\mathrm{e}^{-\ii (H_b + W_s \otimes W_b) \Delta t}\right) \left(\mathrm{e}^{-\ii H_s \Delta t/2} \otimes \id_b\right) \right]^N.
\end{equation}
The next step is to substitute \eqref{eq:Trotter} into \eqref{eq:reduced_density} and evaluate the bath trace analytically. For simplicity, we define
\begin{displaymath}
  \widetilde{\rho}_s(t) = \mathrm{e}^{\ii H_s \Delta t/2} \rho_s(t) \mathrm{e}^{-\ii H_s \Delta t/2},
\end{displaymath}
and the result can be written as
\begin{equation}
  \begin{split}
    \widetilde{\rho}_s(N \Delta t) \approx \sum_{i_1^-, \ldots, i_N^-=1}^M \sum_{i_1^+, \ldots, i_N^+=1}^M
      \mel{i_1^-}{\widetilde{\rho}_{s,0}}{i_1^+} \left( \prod_{k=2}^{N} \mel{i_k^-}{\mathrm{e}^{-\ii H_s \Delta t}}{i_{k-1}^-} \mel{i_{k-1}^+}{\mathrm{e}^{\ii H_s \Delta t}}{i_k^+} \right) \\
      {} \times \mathcal{F}_N(i_1^{\pm}, \ldots, i_N^{\pm}) \ket{i_N^-} \bra{i_N^+},
  \end{split}
  \label{eq:quapi}
\end{equation}
where $\widetilde{\rho}_{s,0} = \mathrm{e}^{-\ii H_s \Delta t/2} \rho_s(0) \mathrm{e}^{\ii H_s \Delta t/2}$, and $\mathcal{F}_N$ is the Feynman-Vernon influence functional, defined as
\begin{equation}
  \mathcal{F}_N(i_1^{\pm}, \ldots, i_N^{\pm}) = \exp\left[-\sum_{j=1}^{N} \sum_{k=1}^{j} (w_{i_j^-} - w_{i_j^+}) (\eta_{j-k} w_{i_k^-} - \eta_{j-k}^* w_{i_k^+})\right],
  \label{eq:influence_functional}
\end{equation}
and the coefficients $\eta_l$ are
\begin{equation}
  \eta_0 = \int_0^{\Delta t} (\Delta t - \tau) B(\tau) \, \mathrm{d}\tau, \qquad
  \eta_l = \int_{-\Delta t}^{\Delta t} (\Delta t - |\tau|) B(l \Delta t + \tau) \, \mathrm{d}\tau, \quad l \geq 1,
  \label{eq:eta}
\end{equation}
with $B(\tau) = Z^{-1}\tr_b[\mathrm{e}^{-\beta H_b} W_b \mathrm{e}^{-\ii \tau H_b} W_b \mathrm{e}^{\ii \tau H_b}]$ being the bath correlation function.

The bath correlation function $B(\tau)$ typically decays rapidly with $\tau$, so that $\eta_l$ also decays with $l$. One can therefore apply the cutoff
\begin{equation}
  \eta_l^{(K_{\max})} = \begin{cases} \eta_l, & 0 \leq l \leq K_{\max}, \\ 0, & l > K_{\max}, \end{cases} \qquad T_{\max} = K_{\max} \Delta t,
  \label{eq:memorytruncation}
\end{equation}
enabling the iterative QuAPI (i-QuAPI) algorithm. We will encounter this approach later in Section~\ref{sec:truncation}, and its details are omitted here.

\subsection{Fast resummation of Dyson series (FRODS)}
\label{sec:frods}
The FRODS method adopts a slightly different splitting of the time-evolution operator:
\begin{equation}
  \begin{split}
    \mathrm{e}^{-\ii H N \Delta t} &= \left(\mathrm{e}^{-\ii H \Delta t}\right)^N   \approx
      \left[ \mathrm{e}^{-\ii H_0 \Delta t/2} \mathrm{e}^{-\ii (W_s \otimes W_b) \Delta t} \mathrm{e}^{-\ii H_0 \Delta t/2} \right]^N \\
    & \approx \left[ \mathrm{e}^{-\ii H_0 \Delta t/2} \left( \mathrm{Id} - \ii \Delta t(W_s \otimes W_b) - \frac{1}{2} \Delta t^2 (W_s \otimes W_b)^2 \right) \mathrm{e}^{-\ii H_0 \Delta t/2} \right]^N,
  \end{split}
  \label{eq:frodssplit}
\end{equation}
where $H_0 = H_s \otimes \id_b + \id_s \otimes H_b$. Define
\begin{equation}
  G_s^{\Delta t} = \mathrm{e}^{-\ii H_s \Delta t}.
\end{equation}
Then substituting \eqref{eq:frodssplit} into \eqref{eq:reduced_density} and evaluating the bath trace gives the following approximation:
\begin{equation}
  \begin{split}
    \widetilde{\rho}_s(N \Delta t) &\approx \sum_{j_{1^-},\ldots, j_{N^-} = 0}^{2}  \sum_{j_{1^+},\ldots, j_{N^+} = 0}^{2} \left( \prod_{k=1}^N \frac{(-\ii \Delta t)^{j_{k^-}}(\ii \Delta t)^{j_{k^+}}}{j_{k^-}! j_{k^+}!} \right) W_s^{j_{N^-}} G_s^{\Delta t} W_s^{j_{(N-1)^-}} G_s^{\Delta t} \cdots G_s^{\Delta t} W_s^{j_{1^-}} \\
    & \qquad {}\times \widetilde{\rho}_{s,0} W_s^{j_{1^+}} (G_s^{\Delta t})^{\dagger} \cdots (G_s^{\Delta t})^{\dagger} W_s^{j_{(N-1)^+}}(G_s^{\Delta t})^{\dagger} W_s^{j_{N^+}} \mathcal{L}_b(j_{N^-}, \ldots, j_{1^-}, j_{1^+}, \ldots, j_{N^+}),
  \end{split}
  \label{eq:frodsapprox}
\end{equation}
where $\mathcal{L}_b$ is obtained by Wick's theorem, whose value is zero if $j_{\mathrm{tot}} := (j_{N^-} + \cdots + j_{1^-}) + (j_{1^+} + \cdots + j_{N^+})$ is odd. When $j_{\mathrm{tot}}$ is even, $\mathcal{L}_b$ is a sum of $(j_{\mathrm{tot}}-1)!!$ terms, each of which is a product of $j_{\mathrm{tot}}/2$ bath correlation functions. The precise form of $\mathcal{L}_b$ is given in Appendix~\ref{app:bath-influence}.

The expansion \eqref{eq:frodsapprox} contains $3^{2N}$ terms in the summation, and each summand requires computing an expensive $\mathcal{L}_b$. To reduce the computational cost, we can follow \cite{sun2026} to perform optimization in the following steps:
\begin{enumerate}
  \item Identify terms that are not needed for the desired order of accuracy. For instance, the first-order method only needs terms with $j_{k^-} + j_{k^+} \leq 1$ for all $k$, while the second-order method requires all terms with
    \begin{equation} \label{eq:m_condition}
      \#\{k : j_{k^-} + j_{k^+} \geqslant 2\} \leq 1 \quad \text{and} \quad j_{k^-} + j_{k^+} \leq 2 \quad \text{for all } k.
    \end{equation}
    For example, when $N = 3$, the following terms can be dropped from the summation without affecting the second-order accuracy:
    \begin{gather*}
      (j_{3^-}, j_{2^-}, j_{1^-}; j_{1^+}, j_{2^+}, j_{3^+}) = (2, 0, 0; 2, 0, 0), \\
      (j_{3^-}, j_{2^-}, j_{1^-}; j_{1^+}, j_{2^+}, j_{3^+}) = (1, 0, 2; 1, 1, 0), \\
      (j_{3^-}, j_{2^-}, j_{1^-}; j_{1^+}, j_{2^+}, j_{3^+}) = (1, 2, 0; 0, 0, 1).
    \end{gather*}
    This can reduce the number of terms in the summation.
  \item Use auxiliary functions $A_n(j_{n^-}, \ldots, j_{1^-}; j_{1^+}, \ldots, j_{n^+})$ (to be defined below) to store partial sums. These auxiliary functions can be computed iteratively, avoiding repetitive calculations.
\end{enumerate} 

For the sake of simplicity, we will only write down the derivations for the first-order method in this section. Define the quantity $A_n(j_{n^-}, \ldots, j_{1^-}; j_{1^+}, \ldots, j_{n^+})$ as
\begin{equation}
  \begin{aligned}
    &A_0(\,;\,) = (G_s^{\Delta t})^{\dagger}\widetilde{\rho}_{s,0}G_s^{\Delta t}, \\
    &A_n(j_{n^-},\ldots,j_{1^-};j_{1^+},\ldots,j_{n^+}) = \sum_{\substack{j'_{1^-} \geqslant j_{1^-}, \, j'_{1^+} \geqslant j_{1^+}\\ j'_{1^-} + j'_{1^+} \leqslant 1}} \cdots \sum_{\substack{j'_{n^-} \geqslant j_{n^-}, \, j'_{n^+} \geqslant j_{n^+}\\ j'_{n^-} + j'_{n^+} \leqslant 1}} \left( \prod_{k=1}^n \frac{(-\ii\Delta t)^{j'_{k^-}} (\ii\Delta t)^{j'_{k^+}}}{j'_{k^-}! j'_{k^+}!}\right) \\
    & \quad \times W_s^{j'_{n^-}} G_s^{\Delta t} W_s^{j'_{(n-1)^-}} G_s^{\Delta t} \cdots G_s^{\Delta t} W_s^{j'_{1^-}} \widetilde{\rho}_{s,0} W_s^{j'_{1^+}} (G_s^{\Delta t})^{\dagger} \cdots (G_s^{\Delta t})^{\dagger} W_s^{j'_{(n-1)^+}} (G_s^{\Delta t})^{\dagger} W_s^{j'_{n^+}} \\
    & \qquad \times \mathcal{L}_b(j'_{n^-} - j_{n^-}, \ldots, j'_{1^-} - j_{1^-}; j'_{1^+} - j_{1^+}, \ldots, j'_{n^+} - j_{n^+}), \qquad j_{k^-} + j_{k^+} \leqslant 1 \text{ for all } k.
  \end{aligned}
  \label{eq:frods-partial-pairings}
\end{equation}
We can observe that $\widetilde{\rho}_s(N\Delta t) \approx A_N(0,\ldots,0;0,\ldots,0)$ by comparing \eqref{eq:frods-partial-pairings} with \eqref{eq:frodsapprox}. For $n\geq1$, the following recurrence relation can be derived directly from the definition \eqref{eq:frods-partial-pairings}:
\begin{equation}
  \begin{aligned}
    &A_n(0,j_{(n-1)^-},\ldots,j_{1^-};j_{1^+},\ldots,j_{(n-1)^+},0) \\
    &\quad=G_s^{\Delta t}A_{n-1}(j_{(n-1)^-},\ldots,j_{1^-};j_{1^+},\ldots,j_{(n-1)^+})(G_s^{\Delta t})^{\dagger} \\
    &\qquad{}-\ii\Delta t\sum_{k=1}^{n-1}B((n-k)\Delta t)
      \Bigl[W_s,G_s^{\Delta t}A_{n-1}(\hat j_{(n-1)^-}^{k^-},\ldots,\hat j_{1^-}^{k^-};\hat j_{1^+}^{k^-},\ldots,\hat j_{(n-1)^+}^{k^-})(G_s^{\Delta t})^{\dagger}\Bigr] \\
    &\qquad{}-\ii\Delta t\sum_{k=1}^{n-1}B((n-k)\Delta t)^*
      \Bigl[W_s,G_s^{\Delta t}A_{n-1}(\hat j_{(n-1)^-}^{k^+},\ldots,\hat j_{1^-}^{k^+};\hat j_{1^+}^{k^+},\ldots,\hat j_{(n-1)^+}^{k^+})(G_s^{\Delta t})^{\dagger}\Bigr].
  \end{aligned}
  \label{eq:frods-zero-update}
\end{equation}
Here $[\cdot,\cdot]$ denotes the commutator and $\hat{j}_{\ell'}^{\ell} = j_{\ell'} + \delta_{\ell \ell'}$. Any $A_{n-1}(j_{(n-1)^-},\ldots,j_{1^-};j_{1^+},\ldots,j_{(n-1)^+})$ is treated as zero if $j_{k^-}+j_{k^+}>1$ for some $k = 1,\ldots,n-1$.

The value of $A_n(j_{n^-}, \ldots, j_{1^-}; j_{1^+}, \ldots, j_{n^+})$ for $j_{n^-} + j_{n^+} = 1$ can also be computed from $A_{n-1}$, and the recurrence relations can again be derived directly from \eqref{eq:frods-partial-pairings}:
\begin{equation}
  \begin{aligned}
    &A_n(1,j_{(n-1)^-},\ldots,j_{1^-};j_{1^+},\ldots,j_{(n-1)^+},0) \\
    &\qquad=-\ii\Delta t\,W_s G_s^{\Delta t}
      A_{n-1}(j_{(n-1)^-},\ldots,j_{1^-};j_{1^+},\ldots,j_{(n-1)^+})(G_s^{\Delta t})^{\dagger}, \\
    &A_n(0,j_{(n-1)^-},\ldots,j_{1^-};j_{1^+},\ldots,j_{(n-1)^+},1) \\
    &\qquad=\ii\Delta t\,G_s^{\Delta t}
      A_{n-1}(j_{(n-1)^-},\ldots,j_{1^-};j_{1^+},\ldots,j_{(n-1)^+})(G_s^{\Delta t})^{\dagger}W_s.
  \end{aligned}
  \label{eq:frods-one-update}
\end{equation}
The equations \eqref{eq:frods-zero-update} and \eqref{eq:frods-one-update} provide a first-order numerical scheme computing $\widetilde{\rho}_s(N\Delta t)$ for any $N$.\footnote{In \cite{sun2026}, an extra term $A_n(1,j_{(n-1)^-},\ldots,j_{1^-};j_{1^+},\ldots,j_{(n-1)^+},1)$ is also included in the recurrence relation. Removing this term does not affect the first-order accuracy of the scheme.}

To show the computational cost, we notice that each pair of $(j_{k^+}, j_{k^-})$ has three admissible values: $(0,0)$, $(1,0)$, and $(0,1)$. Therefore, the total number of auxiliary functions $A_n$ that need to be stored at step $n$ is $3^n$. Each auxiliary function is a matrix of size $M \times M$, so the storage requirement is $O(M^2 3^n)$. For each coefficient at step $n-1$, we compute $G_s^{\Delta t}A_{n-1}(G_s^{\Delta t})^\dagger$ once and reuse it in the updates. This costs $O(M^3)$ per coefficient. Since $W_s$ is diagonal in the chosen basis, multiplication by $W_s$ and evaluation of its commutator with a matrix cost $O(M^2)$. The sums over earlier time steps therefore cost $O(nM^2)$ per output coefficient. The total work at step $n$ is $O(3^n(M^3+nM^2))$, and summing over $N$ steps gives $O(3^N(M^3+NM^2))$. 

The second-order scheme can be derived by including more terms in the definition of $A_n$ \eqref{eq:frods-partial-pairings}, and the recurrence relations are more involved. More details will be provided in a different derivation given in the next section. Here we only mention that the computational cost of the second-order scheme becomes $O(N3^N(M^3+NM^2))$, and the space complexity is $O(M^2 N3^N)$.

\section{Derivation of FRODS from QuAPI}
\label{sec:quapi-to-frods}
This section aims to establish a connection between the QuAPI and FRODS methods. We will show that the FRODS method can be derived from QuAPI with extra approximations. To this end, we extend the expression \eqref{eq:quapi} to the following function:
\begin{equation}
  \begin{split}
    & \mathscr{A}_n(u_{n^-}, \ldots, u_{1^-}; u_{1^+}, \ldots, u_{n^+}) = \\
    & \quad \sum_{i_1^-, \ldots, i_n^-=1}^M \sum_{i_1^+, \ldots, i_n^+=1}^M
      \mel{i_1^-}{\widetilde{\rho}_{s,0}}{i_1^+} \left( \prod_{k=2}^{n} \mel{i_k^-}{G_s^{\Delta t}}{i_{k-1}^-} \mel{i_{k-1}^+}{(G_s^{\Delta t})^{\dagger}}{i_k^+} \right) \\
    & \quad \quad {}\times \exp\left[-\sum_{j=1}^n \sum_{k=1}^{j} (w_{i_j^-} - w_{i_j^+}) (\eta_{j-k} w_{i_k^-} - \eta_{j-k}^* w_{i_k^+}) \underline{{} - \ii \Delta t \sum_{k=1}^n (w_{i_k^-} u_{k^-} - w_{i_k^+} u_{k^+})} \right] \ket{i_n^-} \bra{i_n^+}.
  \end{split}
  \label{eq:A}
\end{equation}
Comparing with \eqref{eq:quapi}, we see that the only difference is the additional underlined term in the exponent, and it is clear that
\begin{displaymath}
  \widetilde{\rho}_s(N\Delta t) = \mathscr{A}_N(0, \ldots, 0; 0, \ldots, 0) + O(\Delta t^2),
\end{displaymath}
where the error $\Delta t^2$ comes from the second-order Trotter splitting. Below, we will find a recursion relation for $\mathscr{A}_n$, which will further lead to an iterative scheme solving $\widetilde{\rho}_s(N\Delta t)$.

Defining
\begin{equation}
  \mathscr{Q}_n(u_{n^-}, \ldots, u_{1^-}; u_{1^+}, \ldots, u_{n^+}) :=
    G_s^{\Delta t}\mathscr{A}_n(u_{n^-}, \ldots, u_{1^-}; u_{1^+}, \ldots, u_{n^+}) (G_s^{\Delta t})^{\dagger},
  \label{eq:Qdef}
\end{equation}
one can find that
\begin{equation}
  \begin{split}
    & \mathscr{Q}_{n-1}(u_{(n-1)^-}, \ldots, u_{1^-}; u_{1^+}, \ldots, u_{(n-1)^+}) \\
    & \quad = \sum_{i_1^-, \ldots, i_n^-=1}^M \sum_{i_1^+, \ldots, i_n^+=1}^M
      \mel{i_1^-}{\widetilde{\rho}_{s,0}}{i_1^+}
      \left( \prod_{k=2}^{n} \mel{i_k^-}{G_s^{\Delta t}}{i_{k-1}^-} \mel{i_{k-1}^+}{(G_s^{\Delta t})^{\dagger}}{i_k^+} \right) \\
    & \qquad {}\times \exp\Biggl[-\sum_{j=1}^{n-1} \sum_{k=1}^{j}
      (w_{i_j^-} - w_{i_j^+}) (\eta_{j-k} w_{i_k^-} - \eta_{j-k}^* w_{i_k^+})
      - \ii \Delta t \sum_{k=1}^{n-1}
      (w_{i_k^-} u_{k^-} - w_{i_k^+} u_{k^+})\Biggr] \ket{i_n^-} \bra{i_n^+}.
  \end{split}
  \label{eq:Q}
\end{equation}
Comparing \eqref{eq:A} and \eqref{eq:Q}, we see that the only difference is that the summation in the exponent of \eqref{eq:Q} is up to $n-1$ instead of $n$. This gives us the following relationship between $\mathscr{A}_n$ and $\mathscr{Q}_{n-1}$:
\begin{equation}
  \begin{split}
    & \mathscr{A}_n(u_{n^-}, \ldots, u_{1^-}; u_{1^+}, \ldots, u_{n^+}) \\
    & \quad = \sum_{i^-,i^+=1}^M
      \mel{i^-}{\mathscr{Q}_{n-1}\!\left(
      u_{(n-1)^-}+c_{(n-1)^-}^{i^-,i^+}, \ldots, u_{1^-}+c_{1^-}^{i^-,i^+};
      u_{1^+}+c_{1^+}^{i^-,i^+}, \ldots, u_{(n-1)^+}+c_{(n-1)^+}^{i^-,i^+}
      \right)}{i^+} \\
    & \hspace{100pt} {}\times \exp\Big(-(w_{i^-}-w_{i^+})(\eta_0 w_{i^-}-\eta_0^* w_{i^+})
      -\ii\Delta t(w_{i^-} u_{n^-}-w_{i^+} u_{n^+})\Big) \dyad{i^-}{i^+},
  \end{split}
  \label{eq:A-Q}
\end{equation}
where, for $1\leq k<n$,
\begin{equation}
  c_{k^-}^{i^-,i^+}=-\frac{\ii}{\Delta t}(w_{i^-}-w_{i^+})\eta_{n-k},
  \qquad
  c_{k^+}^{i^-,i^+}=-\frac{\ii}{\Delta t}(w_{i^-}-w_{i^+})\eta_{n-k}^*.
  \label{eq:ck}
\end{equation}
Note that both $c_{k^-}^{i^-,i^+}$ and $c_{k^+}^{i^-,i^+}$ have order $O(\Delta t)$, and therefore can be regarded as small perturbations of $u_{k^-}$ and $u_{k^+}$, respectively. We can now observe that \eqref{eq:Qdef} and \eqref{eq:A-Q} give a recurrence relation for both $\mathscr{A}_n$ and $\mathscr{Q}_n$.

However, directly using these recurrence relations to find $\widetilde{\rho}_s(N\Delta t) \approx \mathscr{A}_N(0, \ldots, 0; 0, \ldots, 0)$ is still computationally expensive. The time complexity is $O(M^{2N})$, the same as the direct evaluation of \eqref{eq:quapi}. To reduce the computational cost, we will introduce a Taylor expansion to approximate $\mathscr{Q}_{n-1}$ in \eqref{eq:A-Q} and derive an iterative scheme consistent with the FRODS method.

\subsection{Derivation of the first-order method}
To begin with, we consider the first-order Taylor expansion of $\mathscr{A}_n(u_{n^-}, \ldots, u_{1^-}; u_{1^+}, \ldots, u_{n^+})$ given in \eqref{eq:A-Q} in terms of $\Delta t$. Note that $\eta_k = O(\Delta t^2)$ and $c_{k^\pm}^{i^-,i^+} = O(\Delta t)$. Therefore, the first-order expansion gives
\begin{equation}
  \begin{aligned}
    &\mathscr{A}_n(u_{n^-},\ldots,u_{1^-};u_{1^+},\ldots,u_{n^+}) = \\
    &\quad\sum_{i^-,i^+=1}^M\Biggl[
      \bigl(1-\ii\Delta t(w_{i^-}u_{n^-}-w_{i^+}u_{n^+})\bigr)
      \mel{i^-}{\mathscr{Q}_{n-1}(u_{(n-1)^-},\ldots,u_{1^-};u_{1^+},\ldots,u_{(n-1)^+})}{i^+} \\
    &\qquad{}+\sum_{\ell \in \{1^{\pm}, \ldots, (n-1)^{\pm}\}}c_{\ell}^{i^-,i^+}\frac{\partial}{\partial u_{\ell}}
      \mel{i^-}{\mathscr{Q}_{n-1}(u_{(n-1)^-},\ldots,u_{1^-};u_{1^+},\ldots,u_{(n-1)^+})}{i^+} \Biggr]\dyad{i^-}{i^+}+O(\Delta t^2).
  \end{aligned}
  \label{eq:A-Q-first-order}
\end{equation}
To deal with the first-order derivatives in \eqref{eq:A-Q-first-order}, we assume the power series form of $\mathscr{A}_n$ and $\mathscr{Q}_{n-1}$:
\begin{equation}
  \begin{aligned}
    &\mathscr{A}_n(u_{n^-},\ldots,u_{1^-};u_{1^+},\ldots,u_{n^+}) \\
    &\quad=\sum_{j_{n^-},\ldots,j_{1^-}=0}^{\infty}
      \sum_{j_{1^+},\ldots,j_{n^+}=0}^{\infty}
      A_n(j_{n^-},\ldots,j_{1^-};j_{1^+},\ldots,j_{n^+})
      \prod_{k=1}^{n}\frac{u_{k^-}^{j_{k^-}}u_{k^+}^{j_{k^+}}}{j_{k^-}!j_{k^+}!}, \\
    &\mathscr{Q}_{n-1}(u_{(n-1)^-},\ldots,u_{1^-};u_{1^+},\ldots,u_{(n-1)^+}) \\
    &\quad=\sum_{j_{(n-1)^-},\ldots,j_{1^-}=0}^{\infty}
      \sum_{j_{1^+},\ldots,j_{(n-1)^+}=0}^{\infty}
      Q_{n-1}(j_{(n-1)^-},\ldots,j_{1^-};j_{1^+},\ldots,j_{(n-1)^+})
      \prod_{k=1}^{n-1}\frac{u_{k^-}^{j_{k^-}}u_{k^+}^{j_{k^+}}}{j_{k^-}!j_{k^+}!},
  \end{aligned}
  \label{eq:A-Q-power-series}
\end{equation}
where $A_n$ and $Q_{n-1}$ are operator-valued coefficients. It is obvious that
\begin{displaymath}
  \widetilde{\rho}_s(n\Delta t) \approx \mathscr{A}_n(0, \ldots, 0; 0, \ldots, 0) = A_n(0, \ldots, 0; 0, \ldots, 0).
\end{displaymath}
We will now derive a recurrence relation for $A_n$ allowing efficient computation of $\widetilde{\rho}_s(n\Delta t)$.

Plugging \eqref{eq:A-Q-power-series} into \eqref{eq:Qdef} and matching coefficients gives
\begin{equation}
  Q_n(j_{n^-},\ldots,j_{1^-};j_{1^+},\ldots,j_{n^+}) = G_s^{\Delta t} A_n(j_{n^-},\ldots,j_{1^-};j_{1^+},\ldots,j_{n^+}) (G_s^{\Delta t})^{\dagger}.
  \label{eq:Q-A-coeff}
\end{equation}
To explore how to express $A_n$ in terms of $Q_{n-1}$ in a first-order approximation, we define $\hat{j}_{\ell'}^{\ell}=j_{\ell'}+\delta_{\ell'\ell}$. Then, by plugging \eqref{eq:A-Q-power-series} into \eqref{eq:A-Q-first-order} and matching coefficients, we obtain
\begin{equation}
  \begin{aligned}
    & A_n(0,j_{(n-1)^-},\ldots,j_{1^-};j_{1^+},\ldots,j_{(n-1)^+},0) \approx Q_{n-1}(j_{(n-1)^-},\ldots,j_{1^-};j_{1^+},\ldots,j_{(n-1)^+}) \\
    &\qquad{}+\sum_{i^-,i^+=1}^M
      \sum_{\ell\in\{1^{\pm},\ldots,(n-1)^{\pm}\}}c_{\ell}^{i^-,i^+} \mel{i^-}{Q_{n-1}(\hat{j}_{(n-1)^-}^{\ell},\ldots,\hat{j}_{1^-}^{\ell};\hat{j}_{1^+}^{\ell},\ldots,\hat{j}_{(n-1)^+}^{\ell})}{i^+}\dyad{i^-}{i^+}.
  \end{aligned}
  \label{eq:A-coeff-zero}
\end{equation}
Using the expression of $c_{\ell}^{i^-,i^+}$ defined in \eqref{eq:ck}, we can achieve the following more compact form of \eqref{eq:A-coeff-zero}:
\begin{equation}
  \begin{aligned}
    A_n(0,j_{(n-1)^-},\ldots,j_{1^-};j_{1^+},\ldots,j_{(n-1)^+},0) \approx Q_{n-1}(j_{(n-1)^-},\ldots,j_{1^-};j_{1^+},\ldots,j_{(n-1)^+}) \qquad \\
    {}-\frac{\ii}{\Delta t}\sum_{k=1}^{n-1}\eta_{n-k}
      \Bigl[W_s,Q_{n-1}(\hat{j}_{(n-1)^-}^{k^-},\ldots,\hat{j}_{1^-}^{k^-};\hat{j}_{1^+}^{k^-},\ldots,\hat{j}_{(n-1)^+}^{k^-})\Bigr]\phantom{.} \\
    {}-\frac{\ii}{\Delta t}\sum_{k=1}^{n-1}\eta_{n-k}^*
      \Bigl[W_s,Q_{n-1}(\hat{j}_{(n-1)^-}^{k^+},\ldots,\hat{j}_{1^-}^{k^+};\hat{j}_{1^+}^{k^+},\ldots,\hat{j}_{(n-1)^+}^{k^+})\Bigr].
  \end{aligned}
  \label{eq:A-coeff-zero-matrix}
\end{equation}
Here $[\cdot, \cdot]$ denotes the commutator. Similarly, when one of $j_{n^-}$ and $j_{n^+}$ is one and the other is zero, we have
\begin{align}
  A_n(1,j_{(n-1)^-},\ldots,j_{1^-};j_{1^+},\ldots,j_{(n-1)^+},0)
  & \approx-\ii\Delta t\,W_s
    Q_{n-1}(j_{(n-1)^-},\ldots,j_{1^-};j_{1^+},\ldots,j_{(n-1)^+}), \label{eq:A-coeff-one-zero} \\
  A_n(0,j_{(n-1)^-},\ldots,j_{1^-};j_{1^+},\ldots,j_{(n-1)^+},1)
  & = [A_n(1,j_{(n-1)^+},\ldots,j_{1^+};j_{1^-},\ldots,j_{(n-1)^-},0)]^{\dagger}. \label{eq:A-coeff-zero-one}
\end{align}

One can now observe that \eqref{eq:Q-A-coeff} and \eqref{eq:A-coeff-zero-matrix}--\eqref{eq:A-coeff-zero-one} already give us the first-order FRODS iteration defined in \eqref{eq:frods-zero-update}\eqref{eq:frods-one-update}, where we just need to set $Q_n(j_{n-}, \ldots, j_{1-}; j_{1+}, \ldots, j_{n+})$ to zero if $j_{k^-} + j_{k^+} > 1$ for any $1 \leq k \leq n$. The only difference is that the FRODS method introduced in Section~\ref{sec:model} uses $B(k\Delta t)$ instead of $\eta_k/\Delta t^2$ when calculating $A_n(0,j_{(n-1)^-},\ldots,j_{1^-};j_{1^+},\ldots,j_{(n-1)^+},0)$, which does not affect the first-order accuracy as
\begin{displaymath}
  B(k\Delta t) - \frac{\eta_k}{\Delta t^2} = O(\Delta t^2), \qquad k \geqslant 1
\end{displaymath}
according to \eqref{eq:eta}.

\subsection{Derivation of the second-order method}
\label{sec:quapi-to-frods-second-order}
We now claim that the second-order FRODS method can also be derived from \eqref{eq:A-Q} by retaining the quadratic terms in the Taylor expansion. For convenience, we rewrite \eqref{eq:A-Q} as
\begin{equation}
  \begin{split}
    \mathscr{A}_n(u_{n^-},\ldots,u_{1^-};u_{1^+},\ldots,u_{n^+}) =
      \mathscr{Q}_{n-1}(u_{(n-1)^-},\ldots,u_{1^-};u_{1^+},\ldots,u_{(n-1)^+}) \qquad \\
    {} + \mathcal{D}^{\Delta t}_{u} \mathscr{Q}_{n-1}(u_{(n-1)^-},\ldots,u_{1^-};u_{1^+},\ldots,u_{(n-1)^+}) + O(\Delta t^2),
  \end{split}
  \label{eq:A-Q-second-order}
\end{equation}
where $\mathcal{D}^{\Delta t}_{u}$ is the matrix-valued differential operator that can be read off from \eqref{eq:A-Q}:
\begin{equation}
  \mathcal{D}^{\Delta t}_{u}X
  := -\ii\Delta t\bigl(u_{n^-}W_sX-u_{n^+}XW_s\bigr) - \frac{\ii}{\Delta t}\sum_{k=1}^{n-1}
    \left(\eta_{n-k}\left[W_s,\frac{\partial X}{\partial u_{k^-}}\right]
    +\eta_{n-k}^*\left[W_s,\frac{\partial X}{\partial u_{k^+}}\right]\right).
  \label{eq:D-u-def}
\end{equation}
Here $X$ is any differentiable matrix-valued function of the auxiliary variables
$(u_{n^-},\ldots,u_{1^-};u_{1^+},\ldots,u_{n^+})$.

Thus, the second-order expansion of $\mathscr{A}_n$ can be expressed as
\begin{equation}
  \scalebox{.95}{$\begin{split}
    & \mathscr{A}_n(u_{n^-},\ldots,u_{1^-};u_{1^+},\ldots,u_{n^+}) =
      \mathscr{Q}_{n-1}(u_{(n-1)^-},\ldots,u_{1^-};u_{1^+},\ldots,u_{(n-1)^+}) \\
    & \quad {} - \sum_{i^-,i^+=1}^M (w_{i^-}-w_{i^+})(\eta_0 w_{i^-}-\eta_0^* w_{i^+}) \dyad{i^-}{i^-} \mathscr{Q}_{n-1}(u_{(n-1)^-},\ldots,u_{1^-};u_{1^+},\ldots,u_{(n-1)^+}) \dyad{i^+}{i^+} \\
    & \quad {} + \mathcal{D}^{\Delta t}_{u} \left( \mathscr{Q}_{n-1}(u_{(n-1)^-},\ldots,u_{1^-};u_{1^+},\ldots,u_{(n-1)^+}) + \frac{\mathcal{D}^{\Delta t}_{u}}{2} \mathscr{Q}_{n-1}(u_{(n-1)^-},\ldots,u_{1^-};u_{1^+},\ldots,u_{(n-1)^+}) \right) \\
    & \hspace*{430pt} + O(\Delta t^3).
  \end{split}$}
  \label{eq:A-Q-second-order-expanded}
\end{equation}
The term with $\mathcal{D}^{\Delta t}_{u}/2$ in the parentheses is the natural second-order extension of the first-order term $\mathcal{D}^{\Delta t}_{u}$, and the term in the second line comes from the exponential of $-(w_{i^-}-w_{i^+})(\eta_0 w_{i^-}-\eta_0^* w_{i^+})$ in \eqref{eq:A-Q}, which is not involved in the first-order term since $\eta_0 = O(\Delta t^2)$.

Using \eqref{eq:A-Q-second-order-expanded} and \eqref{eq:A-Q-power-series}, we can derive the second-order iterative scheme written in two stages. The first stage computes the intermediate coefficients $A_n^{(1/2)}$ for the quantity in parentheses in \eqref{eq:A-Q-second-order-expanded}:
\begin{equation}
  \begin{aligned}
  &A_n^{(1/2)}(0,j_{(n-1)^-},\ldots,j_{1^-};j_{1^+},\ldots,j_{(n-1)^+},0)=Q_{n-1}(j_{(n-1)^-},\ldots,j_{1^-};j_{1^+},\ldots,j_{(n-1)^+})\\
  &\qquad{}-\frac{\ii}{2\Delta t}\sum_{k=1}^{n-1}\eta_{n-k}
    \Bigl[W_s,Q_{n-1}(\hat{j}_{(n-1)^-}^{k^-},\ldots,\hat{j}_{1^-}^{k^-};\hat{j}_{1^+}^{k^-},\ldots,\hat{j}_{(n-1)^+}^{k^-})\Bigr]\\
  &\qquad{}-\frac{\ii}{2\Delta t}\sum_{k=1}^{n-1}\eta_{n-k}^*
    \Bigl[W_s,Q_{n-1}(\hat{j}_{(n-1)^-}^{k^+},\ldots,\hat{j}_{1^-}^{k^+};\hat{j}_{1^+}^{k^+},\ldots,\hat{j}_{(n-1)^+}^{k^+})\Bigr],\\
  & A_n^{(1/2)}(1,j_{(n-1)^-},\ldots,j_{1^-};j_{1^+},\ldots,j_{(n-1)^+},0)
    =-\frac{\ii\Delta t}{2}W_s Q_{n-1}(j_{(n-1)^-},\ldots,j_{1^-};j_{1^+},\ldots,j_{(n-1)^+}),\\
  & A_n^{(1/2)}(0,j_{(n-1)^-},\ldots,j_{1^-};j_{1^+},\ldots,j_{(n-1)^+},1)
    =[A_n^{(1/2)}(1,j_{(n-1)^+},\ldots,j_{1^+};j_{1^-},\ldots,j_{(n-1)^-},0)]^{\dagger}.
  \end{aligned}
  \label{eq:second-stage-half-zero}
\end{equation}
One can observe that the difference between \eqref{eq:second-stage-half-zero} and \eqref{eq:A-coeff-zero-matrix}--\eqref{eq:A-coeff-zero-one} is merely a factor of $1/2$ for all $O(\Delta t)$ terms, indicating that $A_n^{(1/2)}$ is an intermediate stage analogous to the internal stage of the classical midpoint method. The second stage assembles the final coefficients $A_n$. The three cases needed for the second-order scheme are given below.
\paragraph{Case 1: $\boldsymbol{j_{n^-} = j_{n^+} = 0}$}
\begin{equation}
 \begin{aligned}
 &A_n(0,j_{(n-1)^-},\ldots,j_{1^-};j_{1^+},\ldots,j_{(n-1)^+},0)\approx Q_{n-1}(j_{(n-1)^-},\ldots,j_{1^-};j_{1^+},\ldots,j_{(n-1)^+})\\
 &\qquad{}-\sum_{i^-,i^+=1}^{M}
       (w_{i^-}-w_{i^+})(\eta_0w_{i^-}-\eta_0^*w_{i^+})\dyad{i^-}{i^-}Q_{n-1}(j_{(n-1)^-},\ldots,j_{1^-};j_{1^+},\ldots,j_{(n-1)^+})\dyad{i^+}{i^+}\\
 &\qquad{}-\frac{\ii}{\Delta t}\sum_{k=1}^{n-1}\eta_{n-k}\Bigl[W_s,A_n^{(1/2)}(0,\hat{j}_{(n-1)^-}^{k^-},\ldots,\hat{j}_{1^-}^{k^-};\hat{j}_{1^+}^{k^-},\ldots,\hat{j}_{(n-1)^+}^{k^-},0)\Bigr]\\
 &\qquad{}-\frac{\ii}{\Delta t}\sum_{k=1}^{n-1}\eta_{n-k}^*\Bigl[W_s,A_n^{(1/2)}(0,\hat{j}_{(n-1)^-}^{k^+},\ldots,\hat{j}_{1^-}^{k^+};\hat{j}_{1^+}^{k^+},\ldots,\hat{j}_{(n-1)^+}^{k^+},0)\Bigr].
 \end{aligned}
 \label{eq:second-stage-full}
\end{equation}
\paragraph{Case 2: $\boldsymbol{j_{n^-}+j_{n^+}=1}$}
\begin{equation}
 \begin{aligned}
 &A_n(1,j_{(n-1)^-},\ldots,j_{1^-};j_{1^+},\ldots,j_{(n-1)^+},0)\approx -\ii\Delta t\,W_sA_n^{(1/2)}(0,j_{(n-1)^-},\ldots,j_{1^-};j_{1^+},\ldots,j_{(n-1)^+},0)\\
 &\qquad{}-\frac{\ii}{\Delta t}\sum_{k=1}^{n-1}\eta_{n-k}\Bigl[W_s,A_n^{(1/2)}(1,\hat{j}_{(n-1)^-}^{k^-},\ldots,\hat{j}_{1^-}^{k^-};\hat{j}_{1^+}^{k^-},\ldots,\hat{j}_{(n-1)^+}^{k^-},0)\Bigr]\\
 &\qquad{}-\frac{\ii}{\Delta t}\sum_{k=1}^{n-1}\eta_{n-k}^*\Bigl[W_s,A_n^{(1/2)}(1,\hat{j}_{(n-1)^-}^{k^+},\ldots,\hat{j}_{1^-}^{k^+};\hat{j}_{1^+}^{k^+},\ldots,\hat{j}_{(n-1)^+}^{k^+},0)\Bigr],\\
 &A_n(0,j_{(n-1)^-},\ldots,j_{1^-};j_{1^+},\ldots,j_{(n-1)^+},1) = [A_n(1,j_{(n-1)^+},\ldots,j_{1^+};j_{1^-},\ldots,j_{(n-1)^-},0)]^{\dagger}.
 \end{aligned}
 \label{eq:second-stage-full-minus}
\end{equation}
\paragraph{Case 3: $\boldsymbol{j_{n^-}+j_{n^+}=2}$}
\begin{equation}
  \begin{aligned}
    A_n(2,j_{(n-1)^-},\ldots,j_{1^-};j_{1^+},\ldots,j_{(n-1)^+},0)&\approx -2\ii\Delta t\,W_sA_n^{(1/2)}(1,j_{(n-1)^-},\ldots,j_{1^-};j_{1^+},\ldots,j_{(n-1)^+},0),\\
    A_n(0,j_{(n-1)^-},\ldots,j_{1^-};j_{1^+},\ldots,j_{(n-1)^+},2)&=[A_n(2,j_{(n-1)^+},\ldots,j_{1^+};j_{1^-},\ldots,j_{(n-1)^-},0)]^{\dagger}, \\
    A_n(1,j_{(n-1)^-},\ldots,j_{1^-};j_{1^+},\ldots,j_{(n-1)^+},1)&\approx -\ii\Delta t\,W_sA_n^{(1/2)}(0,j_{(n-1)^-},\ldots,j_{1^-};j_{1^+},\ldots,j_{(n-1)^+},1)\\
    &\quad{}+\ii\Delta t\,A_n^{(1/2)}(1,j_{(n-1)^-},\ldots,j_{1^-};j_{1^+},\ldots,j_{(n-1)^+},0)W_s.
  \end{aligned}
 \label{eq:second-stage}
\end{equation}
Equations \eqref{eq:Q-A-coeff} and \eqref{eq:second-stage-half-zero}--\eqref{eq:second-stage} together give the second-order FRODS iteration. As discussed in the previous section, we are only interested in $A_n(j_{n-}, \ldots, j_{1-}; j_{1+}, \ldots, j_{n+})$ satisfying \eqref{eq:m_condition}, and when computing the intermediate coefficients $A_n^{(1/2)}$, only quantities of the form \eqref{eq:second-stage-half-zero} are needed, as the first stage of the midpoint method requires only first order of accuracy. Any quantity with indices violating such constraints should be regarded as zero in the computation.

Note that this numerical scheme does not exactly reproduce the sum \eqref{eq:frodsapprox} with coefficients constrained by \eqref{eq:m_condition}, but the approximation of the coefficient $A_N(0, \ldots, 0; 0, \ldots, 0)$ computed by the scheme covers all such terms with suitable replacements of bath correlation functions, and the second-order accuracy is still guaranteed.

We can now look back at the time and space complexity claimed at the end of Section~\ref{sec:model}. Under the restriction \eqref{eq:m_condition}, either every pair $(j_{k^-},j_{k^+})$ has one of the three values $(0,0)$, $(1,0)$, and $(0,1)$, or exactly one pair has one of the additional values $(2,0)$, $(1,1)$, and $(0,2)$. The first case gives $3^n$ auxiliary functions, while the second gives $n\cdot3\cdot3^{n-1}$. Thus, the total number is $(n+1)3^n$, and storing these $M\times M$ matrices requires $O(M^2 n3^n)$ space. The intermediate coefficients require at most the same order of storage. As in Section~\ref{sec:model}, computing each transferred coefficient $Q_{n-1}$ costs $O(M^3)$, and these coefficients are reused in both Taylor stages. Each stage costs $O(nM^2)$ per output coefficient because $W_s$ is diagonal and the sums contain $O(n)$ terms. The self-interaction correction in~\eqref{eq:second-stage-full} acts entrywise and costs $O(M^2)$ per coefficient. The intermediate coefficients are computed once and reused in the second stage. Consequently, the cost at step $n$ is $O(n3^n(M^3+nM^2))$, and summing over $N$ steps gives $O(N3^N(M^3+NM^2))$ total work and $O(M^2 N3^N)$ peak storage.

In the derivations below, we will follow the same approach to obtain new iterative schemes in four steps:
\begin{enumerate}
  \item Define the transferred auxiliary function $\mathscr{Q}_{n-1}$ in terms of $\mathscr{A}_{n-1}$.
  \item Express $\mathscr{A}_n$ in terms of $\mathscr{Q}_{n-1}$ through an exact recurrence.
  \item Expand this recurrence in $\Delta t$ to the desired order.
  \item Substitute the Taylor series in the auxiliary variables $u_{k^\pm}$ and match coefficients to obtain the updates for $A_n$ and $Q_{n-1}$.
\end{enumerate}
The main change will be the definition of $\mathscr{Q}_{n-1}$. By incorporating the near-field bath influence factors into this function and retaining them in exponential form, we avoid their Taylor truncation, which can improve accuracy. The remaining far-field contribution is still treated by expansion in $\Delta t$. We therefore refer to the resulting schemes as path integral with far-field expansion (PIFFE) methods, which we will describe in detail in the next section.

\section{Path integral with far-field expansion (PIFFE)}
\label{sec:localresummation}
Compared to QuAPI, the FRODS method reduces computational and storage costs by applying a Taylor expansion to the bath influence functional (see \eqref{eq:A-Q-first-order} and \eqref{eq:A-Q-second-order-expanded}). However, as we will show in Section~\ref{sec:numerics}, the Taylor expansion can introduce significant errors when the coupling between the system and bath is strong. In this section, we introduce the path integral method with far-field expansion (PIFFE). We retain interactions spanning at most $r$ time steps in exponential form and apply a Taylor expansion to the remaining contribution. Here, the far field refers to interactions between time steps $j$ and $k$ with $j-k>r$. We first demonstrate the idea in Section~\ref{sec:selfinteraction} and present the full algorithm in Section~\ref{sec:generalradius}.

\subsection{A motivating example}
\label{sec:selfinteraction}
The first-order Taylor expansion of the bath influence functional essentially replaces $\mathrm{e}^x$ with $1+x$ for small $x$. In FRODS, the variable $x$ refers to $\mathcal{F}_{jk} = -(w_{i_j^-} - w_{i_j^+})(\eta_{j-k} w_{i_k^-} - \eta_{j-k}^* w_{i_k^+})$. When $|\eta_{j-k}|$ is large, this approximation can introduce significant errors. Since the bath correlation function $B(t)$ decays with time, $|\eta_{j-k}|$ may be large when $j$ is close to $k$. Motivated by this observation, we aim to retain the exponential form when $j-k$ is small. In this subsection, we focus on a simple case in which only the factor $\mathcal{F}_{jk}$ with $j = k$ is retained in exponential form.

To this end, we slightly modify the definition of $\mathscr{Q}_{n-1}$ in \eqref{eq:A-Q} to include the factor $\mathrm{e}^{\mathcal{F}_{nn}}$:
\begin{equation}
  \begin{aligned}
    &\mathscr{Q}^{(0)}_{n-1}(u_{(n-1)^-},\ldots,u_{1^-};u_{1^+},\ldots,u_{(n-1)^+}) \\
    &\quad = \mathcal{E}^{(0)}\!\left[
      G_s^{\Delta t} \mathscr{A}_{n-1}(u_{(n-1)^-},\ldots,u_{1^-};u_{1^+},\ldots,u_{(n-1)^+}) (G_s^{\Delta t})^{\dagger}\right],
  \end{aligned}
  \label{eq:Qdef-r0}
\end{equation}
where the superoperator $\mathcal{E}^{(0)}$ acts on a system matrix $X$ as
\begin{equation}
  \mathcal{E}^{(0)}[X] = \sum_{i^-,i^+=1}^{M} e^{-(w_{i^-}-w_{i^+})(\eta_0 w_{i^-}-\eta_0^* w_{i^+})} \dyad{i^-}{i^-} X \dyad{i^+}{i^+}.
\end{equation}
With this definition, the recurrence relation \eqref{eq:A-Q} becomes
\begin{equation}
  \begin{split}
    & \mathscr{A}_n(u_{n^-}, \ldots, u_{1^-}; u_{1^+}, \ldots, u_{n^+}) \\
    & \quad = \sum_{i^-,i^+=1}^M
      \mel{i^-}{\mathscr{Q}^{(0)}_{n-1}\!\left(
      u_{(n-1)^-}+c_{(n-1)^-}^{i^-,i^+}, \ldots, u_{1^-}+c_{1^-}^{i^-,i^+};
      u_{1^+}+c_{1^+}^{i^-,i^+}, \ldots, u_{(n-1)^+}+c_{(n-1)^+}^{i^-,i^+}
      \right)}{i^+} \\
    & \hspace{250pt} {}\times \exp\Big(-\ii\Delta t(w_{i^-} u_{n^-}-w_{i^+} u_{n^+})\Big) \dyad{i^-}{i^+}.
  \end{split}
  \label{eq:A-Q-r0}
\end{equation}
Equations \eqref{eq:Qdef-r0} and \eqref{eq:A-Q-r0} define an exact recurrence relation for $\mathscr{A}_n$, from which we are able to derive a numerical scheme that retains the factor $\mathrm{e}^{\mathcal{F}_{nn}}$ exactly.

The derivation follows the same steps as in Section~\ref{sec:quapi-to-frods}. We first apply a Taylor expansion to \eqref{eq:A-Q-r0} and obtain a result similar to \eqref{eq:A-Q-second-order-expanded}:
\begin{equation}
  \scalebox{.95}{$\begin{split}
    & \mathscr{A}_n(u_{n^-},\ldots,u_{1^-};u_{1^+},\ldots,u_{n^+}) =
      \mathscr{Q}_{n-1}^{(0)}(u_{(n-1)^-},\ldots,u_{1^-};u_{1^+},\ldots,u_{(n-1)^+}) \\
    & \quad {} + \mathcal{D}^{\Delta t}_{u} \left( \mathscr{Q}_{n-1}^{(0)}(u_{(n-1)^-},\ldots,u_{1^-};u_{1^+},\ldots,u_{(n-1)^+}) + \frac{\mathcal{D}^{\Delta t}_{u}}{2} \mathscr{Q}_{n-1}^{(0)}(u_{(n-1)^-},\ldots,u_{1^-};u_{1^+},\ldots,u_{(n-1)^+}) \right) \\
    & \hspace*{430pt} + O(\Delta t^3).
  \end{split}$}
  \label{eq:A-Q-r0-second-order-expanded}
\end{equation}
The only difference from \eqref{eq:A-Q-second-order-expanded} is that the second line in \eqref{eq:A-Q-second-order-expanded} is removed because we have already included the exponential term in \eqref{eq:Qdef-r0}. We then consider the power series expansions of both $\mathscr{A}_n$ and $\mathscr{Q}^{(0)}_{n-1}$. The operator $\mathscr{A}_n$ is still expanded as in \eqref{eq:A-Q-power-series}, while $\mathscr{Q}^{(0)}_{n-1}$ is expanded as
\begin{equation}
  \begin{aligned}
    &\mathscr{Q}^{(0)}_{n-1}(u_{(n-1)^-},\ldots,u_{1^-};u_{1^+},\ldots,u_{(n-1)^+}) \\
    &\quad = \sum_{j_{(n-1)^-},\ldots,j_{1^-}=0}^{\infty}
      \sum_{j_{1^+},\ldots,j_{(n-1)^+}=0}^{\infty}
      Q^{(0)}_{n-1}(j_{(n-1)^-},\ldots,j_{1^-};j_{1^+},\ldots,j_{(n-1)^+})
      \prod_{k=1}^{n-1}\frac{u_{k^-}^{j_{k^-}}u_{k^+}^{j_{k^+}}}{j_{k^-}!j_{k^+}!}.
  \end{aligned}
  \label{eq:Q-power-series}
\end{equation}
Substituting the expansions into \eqref{eq:Qdef-r0} yields
\begin{equation}
  \begin{aligned}
    &Q^{(0)}_{n-1}(j_{(n-1)^-},\ldots,j_{1^-};j_{1^+},\ldots,j_{(n-1)^+}) \\
    &\quad = \mathcal{E}^{(0)}\!\left[
      G_s^{\Delta t}A_{n-1}(j_{(n-1)^-},\ldots,j_{1^-};j_{1^+},\ldots,j_{(n-1)^+}) (G_s^{\Delta t})^{\dagger}\right].
  \end{aligned}
  \label{eq:Q-coeff-r0}
\end{equation}
The recurrence relation expressing $A_n$ in terms of $Q^{(0)}_{n-1}$ is obtained by applying the power series expansion to \eqref{eq:A-Q-r0} and matching coefficients. To obtain the first-order scheme, the term with $\mathcal{D}^{\Delta t}_{u}/2$ in \eqref{eq:A-Q-r0-second-order-expanded} can be dropped, and the resulting recurrence relation is the same as \eqref{eq:A-coeff-zero-matrix}--\eqref{eq:A-coeff-zero-one}, except that $Q_{n-1}$ is replaced by $Q^{(0)}_{n-1}$. Again, all quantities with $j_{k^-}+j_{k^+}>1$ for any $1\le k\le n$ are set to zero.

To obtain the second-order scheme, we follow the same steps as in Section~\ref{sec:quapi-to-frods} to derive the two-stage iteration. The first stage again computes the intermediate coefficients $A_n^{(1/2)}$, corresponding to the Taylor coefficients of the quantity in parentheses in \eqref{eq:A-Q-r0-second-order-expanded}, using the same procedure as in \eqref{eq:second-stage-half-zero}, except that $Q_{n-1}$ is replaced by $Q^{(0)}_{n-1}$. The second stage computes $A_n$ from $A_n^{(1/2)}$, and the only differences from \eqref{eq:second-stage-full}--\eqref{eq:second-stage} are that $Q_{n-1}$ is replaced by $Q^{(0)}_{n-1}$ and that the second line in \eqref{eq:second-stage-full} is removed, since the second line of \eqref{eq:A-Q-second-order-expanded} is no longer present in \eqref{eq:A-Q-r0-second-order-expanded}. This gives the following update for $A_n(0,j_{(n-1)^-},\ldots,j_{1^-};j_{1^+},\ldots,j_{(n-1)^+},0)$: 
\begin{equation}
  \begin{aligned}
  A_n(0,j_{(n-1)^-},\ldots,j_{1^-};j_{1^+},\ldots,j_{(n-1)^+},0)\approx Q^{(0)}_{n-1}(j_{(n-1)^-},\ldots,j_{1^-};j_{1^+},\ldots,j_{(n-1)^+})\\
  -\frac{\ii}{\Delta t}\sum_{k=1}^{n-1}\eta_{n-k}\Bigl[W_s,A_n^{(1/2)}(0,\hat{j}_{(n-1)^-}^{k^-},\ldots,\hat{j}_{1^-}^{k^-};\hat{j}_{1^+}^{k^-},\ldots,\hat{j}_{(n-1)^+}^{k^-},0)\Bigr]\\
  -\frac{\ii}{\Delta t}\sum_{k=1}^{n-1}\eta_{n-k}^*\Bigl[W_s,A_n^{(1/2)}(0,\hat{j}_{(n-1)^-}^{k^+},\ldots,\hat{j}_{1^-}^{k^+};\hat{j}_{1^+}^{k^+},\ldots,\hat{j}_{(n-1)^+}^{k^+},0)\Bigr].
  \end{aligned}
  \label{eq:r0-second-stage-full}
\end{equation}
Equations \eqref{eq:second-stage-full-minus} and \eqref{eq:second-stage} remain unchanged. During the iteration, we retain only $A_n(j_{n^-}, \ldots, j_{1^-}; j_{1^+}, \ldots, j_{n^+})$ satisfying \eqref{eq:m_condition}, and treat any quantity whose indices violate these constraints as zero in the computation.

\subsection{Derivation of the general PIFFE scheme}
\label{sec:generalradius}
In the previous subsection, we redefined $\mathscr{Q}$ to include a potentially large factor $\mathrm{e}^{\mathcal{F}_{nn}}$ appearing in the definition of $\mathscr{A}$, so that the Taylor expansion of $\mathscr{A}$ retains the exponential form of this factor. As discussed above, $\mathcal{F}_{jk} = -(w_{i_j^-} - w_{i_j^+})(\eta_{j-k} w_{i_k^-} - \eta_{j-k}^* w_{i_k^+})$ could be large when $j-k$ is small, which motivates us to include all $\mathcal{F}_{nk}$ with $0\le n-k\le r$ in the definition of $\mathscr{Q}$. The factor involving all such $\mathcal{F}_{nk}$ is
\begin{equation}
  \exp\!\left[-(w_{i_n^-}-w_{i_n^+})
  \sum_{l=0}^{\min(r,n-1)}
  (\eta_lw_{i_{n-l}^-}-\eta_l^*w_{i_{n-l}^+})\right].
  \label{eq:derivedpathweight}
\end{equation}
Since this factor depends not only on $(i_n^-,i_n^+)$, but also on previous indices $(i_{n-1}^-,i_{n-1}^+),\ldots,(i_{n-\tilde{r}}^-,i_{n-\tilde{r}}^+)$, where $\tilde{r} = \min(r,n-1)$, we can no longer define $\mathscr{Q}$ to be an $M\times M$ matrix as before. Instead, we let both $\mathscr{A}_n^{(\tilde r)}$ and $\mathscr{Q}_{n-1}^{(\tilde r)}$ be tensors whose entries are given by $\mel{i_n^-,\ldots,i_{n-\tilde{r}}^-}{\mathscr{A}_n^{(\tilde r)}}{i_n^+,\ldots,i_{n-\tilde{r}}^+}$ and $\mel{i_n^-,\ldots,i_{n-\tilde{r}}^-}{\mathscr{Q}_{n-1}^{(\tilde r)}}{i_n^+,\ldots,i_{n-\tilde{r}}^+}$, respectively. These tensors can also be regarded as operators acting on $(\mathbb C^M)^{\otimes(\tilde r+1)}$. The precise definition of $\mathscr{A}_n^{(\tilde r)}$ for $n \geqslant 1$ is
\begin{equation}
  \begin{aligned}
    &\mathscr{A}_n^{(\tilde r)}(u_{n^-},\ldots,u_{1^-};u_{1^+},\ldots,u_{n^+}) \\
    &\quad=\sum_{i_1^-,\ldots,i_n^-=1}^{M}\sum_{i_1^+,\ldots,i_n^+=1}^{M}
      \mel{i_1^-}{\widetilde{\rho}_{s,0}}{i_1^+}
      \left(\prod_{k=2}^{n}\mel{i_k^-}{G_s^{\Delta t}}{i_{k-1}^-}
      \mel{i_{k-1}^+}{(G_s^{\Delta t})^{\dagger}}{i_k^+}\right) \\
    &\qquad{}\times\exp\Biggl[-\sum_{j=1}^{n}\sum_{k=1}^{j}
      (w_{i_j^-}-w_{i_j^+})(\eta_{j-k}w_{i_k^-}-\eta_{j-k}^*w_{i_k^+})
      -\ii\Delta t\sum_{k=1}^{n}(w_{i_k^-}u_{k^-}-w_{i_k^+}u_{k^+})\Biggr] \\
    &\qquad{}\times\dyad{i_n^-,\ldots,i_{n-\tilde r}^-}{i_n^+,\ldots,i_{n-\tilde r}^+}.
  \end{aligned}
\end{equation}
Compared to \eqref{eq:A}, the only difference is that the dyad in the last line now contains $\tilde r+1$ indices instead of just one. The definition of $\mathscr{Q}_{n-1}^{(\tilde r)}$ is an extension of \eqref{eq:Qdef-r0}:
\begin{equation}
 \begin{aligned}
 &\mathscr Q_{n-1}^{(\tilde{r})}(u_{(n-1)^-},\ldots,u_{1^-};u_{1^+},\ldots,u_{(n-1)^+})\\
 &\quad=\begin{cases}
 \mathcal E^{(n-1)}\!\Bigl[\widetilde{V}^{(n-1)}\mathscr A_{n-1}^{(n-2)}(u_{(n-1)^-},\ldots,u_{1^-};u_{1^+},\ldots,u_{(n-1)^+}) (\widetilde{V}^{(n-1)})^\dagger\Bigr],
 & 2\le n\le r+1,\\[6pt]
 \mathcal E^{(r)}\!\Bigl[V^{(r)}\mathscr A_{n-1}^{(r)}(u_{(n-1)^-},\ldots,u_{1^-};u_{1^+},\ldots,u_{(n-1)^+}) (V^{(r)})^\dagger\Bigr],
 & n\ge r+2.
 \end{cases}
 \end{aligned}
 \label{eq:derivedtransfer}
\end{equation}
The superoperator $\mathcal E^{(\tilde{r})}$ is defined as
\begin{equation}
  \begin{aligned}
  \mathcal E^{(\tilde{r})}[X]
  &=\sum_{i_0^-,\ldots,i_{\tilde{r}}^-=1}^{M}\sum_{i_0^+,\ldots,i_{\tilde{r}}^+=1}^{M}
  \exp\!\left[-(w_{i_0^-}-w_{i_0^+})
  \sum_{l=0}^{\tilde r}(\eta_lw_{i_l^-}-\eta_l^*w_{i_l^+})\right]\\
  &\qquad{}\times\dyad{i_0^-,\ldots,i_{\tilde{r}}^-}{i_0^-,\ldots,i_{\tilde{r}}^-}
  X\dyad{i_0^+,\ldots,i_{\tilde{r}}^+}{i_0^+,\ldots,i_{\tilde{r}}^+},
  \end{aligned}
  \label{eq:generalnearmap}
\end{equation}
and the two operators $\widetilde{V}^{(n-1)}$ and $V^{(r)}$ are defined by
\begin{equation}
 \begin{aligned}
  \widetilde{V}^{(n-1)}: & (\mathbb{C}^M)^{\otimes (n-1)} \to (\mathbb{C}^M)^{\otimes n}, &
 \widetilde{V}^{(n-1)}\ket{i_0,\ldots,i_{n-2}}
 &=\sum_{i=1}^{M}\mel{i}{G_s^{\Delta t}}{i_0}
   \ket{i,i_0,\ldots,i_{n-2}},\\
  V^{(r)}: & (\mathbb{C}^M)^{\otimes (r+1)} \to (\mathbb{C}^M)^{\otimes (r+1)}, &
 V^{(r)}\ket{i_0,\ldots,i_r}
 &=\sum_{i=1}^{M}\mel{i}{G_s^{\Delta t}}{i_0}
   \ket{i,i_0,\ldots,i_{r-1}}.
 \end{aligned}
 \label{eq:historytransfermatrix}
\end{equation}
Note that $\widetilde{V}^{(n-1)}$ increases the tensor order by one, thereby extending the history, while $V^{(r)}$ keeps the tensor order unchanged, so that the tensor order of $\mathscr{Q}_{n-1}^{(\tilde r)}$ matches that of $\mathscr{A}_n^{(\tilde r)}$. For $r=0$, we have $V^{(0)}=G_s^{\Delta t}$, and therefore the definition of $\mathscr{Q}_{n-1}^{(0)}$ in \eqref{eq:derivedtransfer} reduces to \eqref{eq:Qdef-r0}.

As in \eqref{eq:A-Q-r0}, one can verify by straightforward calculations that $\mathscr{A}_n^{(\tilde r)}$ can be expressed in terms of $\mathscr{Q}_{n-1}^{(\tilde r)}$ as
\begin{equation}
 \begin{aligned}
 &\mathscr A_n^{(\tilde r)}(u_{n^-},\ldots,u_{1^-};u_{1^+},\ldots,u_{n^+})\\
 &\quad=\sum_{i_{n-\tilde r}^-,\ldots,i_n^-=1}^{M}
 \sum_{i_{n-\tilde r}^+,\ldots,i_n^+=1}^{M}
 \mel{i_n^-,\ldots,i_{n-\tilde r}^-}{\mathscr{Q}_{n-1}^{(\tilde r)}(\hat{u}_{(n-1)^-},\ldots,\hat{u}_{1^-};\hat{u}_{1^+},\ldots,\hat{u}_{(n-1)^+})}{i_n^+,\ldots,i_{n-\tilde r}^+}\\
 & \qquad {} \times \exp\left[-\ii\Delta t \left(w_{i_n^-}u_{n^-}-w_{i_n^+}u_{n^+}\right) \right]\dyad{i_n^-,\ldots,i_{n-\tilde r}^-}{i_n^+,\ldots,i_{n-\tilde r}^+},
 \end{aligned}
 \label{eq:derivedtransfermatrix}
\end{equation}
where
\begin{equation}
 \begin{aligned}
 \hat{u}_{k^-}
 &=\begin{cases}
 u_{k^-}-\dfrac{\ii}{\Delta t}(w_{i_n^-}-w_{i_n^+})\eta_{n-k},&n-k>r,\\
 u_{k^-},&n-k\le r,
 \end{cases}\\[4pt]
 \hat{u}_{k^+}
 &=\begin{cases}
 u_{k^+}-\dfrac{\ii}{\Delta t}(w_{i_n^-}-w_{i_n^+})\eta_{n-k}^*,&n-k>r,\\
 u_{k^+},&n-k\le r.
 \end{cases}
 \end{aligned}
\end{equation}
Note that $\hat{u}_{k^{\pm}} = u_{k^{\pm}}$ for $n-k\le r$, since the near interactions, expressed as exponentials in \eqref{eq:generalnearmap}, are already included in $\mathscr Q_{n-1}^{(\tilde r)}$. Equations \eqref{eq:derivedtransfer} and \eqref{eq:derivedtransfermatrix} define a recurrence relation for $\mathscr{A}_n^{(\tilde r)}$, and the reduced density matrix can be obtained by
\begin{equation}
 \begin{aligned}
 \widetilde{\rho}_s(n\Delta t)
 &\approx\sum_{i_{n-\tilde r}^-,\ldots,i_n^-=1}^{M}
   \sum_{i_{n-\tilde r}^+,\ldots,i_n^+=1}^{M}
   \mel{i_n^-,\ldots,i_{n-\tilde r}^-}
   {\mathscr{A}_n^{(\tilde r)}(0,\ldots,0;0,\ldots,0)}
   {i_n^+,\ldots,i_{n-\tilde r}^+} \dyad{i_n^-}{i_n^+}.
 \end{aligned}
\end{equation}

We now expand \eqref{eq:derivedtransfermatrix} to second order in $\Delta t$. Following \eqref{eq:D-u-def}, we define the first-order differential operator $\mathcal D_u^{\Delta t,(\tilde r)}$:
\begin{equation}
  \mathcal D_u^{\Delta t,(\tilde r)}X
  =-\ii\Delta t\left(u_{n^-}W^{(\tilde r)}X-u_{n^+}XW^{(\tilde r)}\right)
  -\frac{\ii}{\Delta t}
  \sum_{k=1}^{n-\tilde r-1}
  \left[W^{(\tilde r)},\eta_{n-k}\frac{\partial X}{\partial u_{k^-}}
  +\eta_{n-k}^*\frac{\partial X}{\partial u_{k^+}}\right],
  \label{eq:generaldifferentialaction}
\end{equation}
where $X$ is an operator acting on $(\mathbb C^M)^{\otimes(\tilde r+1)}$ with parameters $u_{n^-},\ldots,u_{1^-};u_{1^+},\ldots,u_{n^+}$, and
\begin{equation}
  W^{(\tilde r)}
  =\sum_{i_{n-\tilde r},\ldots,i_n=1}^{M}
  w_{i_n}\dyad{i_n,\ldots,i_{n-\tilde r}}{i_n,\ldots,i_{n-\tilde r}}.
\end{equation}
Thus, the quadratic Taylor approximation is
\begin{equation}
  \mathscr A_n^{(\tilde r)}\approx\mathscr Q_{n-1}^{(\tilde r)}
  +\mathcal D_u^{\Delta t,(\tilde r)}\left(
  \mathscr Q_{n-1}^{(\tilde r)}
  +\tfrac12\mathcal D_u^{\Delta t,(\tilde r)}\mathscr Q_{n-1}^{(\tilde r)}\right).
  \label{eq:generalquadraticexpansion}
\end{equation}
We next substitute the power series expansions of both $\mathscr A_n^{(\tilde r)}$ and $\mathscr Q_{n-1}^{(\tilde r)}$ into \eqref{eq:derivedtransfer} and \eqref{eq:generalquadraticexpansion} to relate their coefficients.

As in~\eqref{eq:A-Q-power-series}, we use the factorial expansions
\begin{equation}
 \begin{aligned}
 &\mathscr A_n^{(\tilde r)}(u_{n^-},\ldots,u_{1^-};u_{1^+},\ldots,u_{n^+})\\
 &\quad=\sum_{j_{n^-},\ldots,j_{1^-}=0}^{\infty}
 \sum_{j_{1^+},\ldots,j_{n^+}=0}^{\infty}
 A_n^{(\tilde r)}(j_{n^-},\ldots,j_{1^-};j_{1^+},\ldots,j_{n^+})
 \prod_{k=1}^{n}\frac{u_{k^-}^{j_{k^-}}u_{k^+}^{j_{k^+}}}{j_{k^-}!j_{k^+}!},\\
 &\mathscr Q_{n-1}^{(\tilde r)}(u_{(n-1)^-},\ldots,u_{1^-};u_{1^+},\ldots,u_{(n-1)^+})\\
 &\quad=\sum_{j_{(n-1)^-},\ldots,j_{1^-}=0}^{\infty}
 \sum_{j_{1^+},\ldots,j_{(n-1)^+}=0}^{\infty}
 Q_{n-1}^{(\tilde r)}(j_{(n-1)^-},\ldots,j_{1^-};j_{1^+},\ldots,j_{(n-1)^+})
 \prod_{k=1}^{n-1}\frac{u_{k^-}^{j_{k^-}}u_{k^+}^{j_{k^+}}}{j_{k^-}!j_{k^+}!}.
 \end{aligned}
 \label{eq:generalcoefficients}
\end{equation}
Applying the power series and matching coefficients in \eqref{eq:derivedtransfer} gives
\begin{equation}
 \begin{aligned}
 &Q_{n-1}^{(\tilde r)}(j_{(n-1)^-},\ldots,j_{1^-};j_{1^+},\ldots,j_{(n-1)^+})\\
 &\quad=\begin{cases}
 \mathcal E^{(n-1)}\!\Bigl[\widetilde V^{(n-1)}A_{n-1}^{(n-2)}(j_{(n-1)^-},\ldots,j_{1^-};j_{1^+},\ldots,j_{(n-1)^+})
 (\widetilde V^{(n-1)})^\dagger\Bigr],
 &2\le n\le r+1,\\[6pt]
 \mathcal E^{(r)}\!\Bigl[V^{(r)}A_{n-1}^{(r)}(j_{(n-1)^-},\ldots,j_{1^-};j_{1^+},\ldots,j_{(n-1)^+})
 (V^{(r)})^\dagger\Bigr],
 &n\ge r+2.
 \end{cases}
 \end{aligned}
 \label{eq:generaltransfer}
\end{equation}
Matching coefficients in \eqref{eq:generalquadraticexpansion} gives the update rule for $A_n^{(\tilde r)}$ in terms of $Q_{n-1}^{(\tilde r)}$. As before, we retain only $A_n^{(\tilde r)}(j_{n^-},\ldots,j_{1^-};j_{1^+},\ldots,j_{n^+})$ satisfying \eqref{eq:m_condition}. The intermediate coefficients, analogous to those in \eqref{eq:second-stage-half-zero}, are given by
\begin{equation}
 \begin{aligned}
 &A_n^{(1/2)}(0,j_{(n-1)^-},\ldots,j_{1^-};j_{1^+},\ldots,j_{(n-1)^+},0)=Q_{n-1}^{(\tilde r)}(j_{(n-1)^-},\ldots,j_{1^-};j_{1^+},\ldots,j_{(n-1)^+})\\
 &\qquad{}-\frac{\ii}{2\Delta t}\sum_{k=1}^{n-\tilde r-1}\eta_{n-k}
 \Bigl[W^{(\tilde r)},Q_{n-1}^{(\tilde r)}
 (\hat j_{(n-1)^-}^{k^-},\ldots,\hat j_{1^-}^{k^-};
  \hat j_{1^+}^{k^-},\ldots,\hat j_{(n-1)^+}^{k^-})\Bigr]\\
 &\qquad{}-\frac{\ii}{2\Delta t}\sum_{k=1}^{n-\tilde r-1}\eta_{n-k}^*
 \Bigl[W^{(\tilde r)},Q_{n-1}^{(\tilde r)}
 (\hat j_{(n-1)^-}^{k^+},\ldots,\hat j_{1^-}^{k^+};
  \hat j_{1^+}^{k^+},\ldots,\hat j_{(n-1)^+}^{k^+})\Bigr],\\
 &A_n^{(1/2)}(1,j_{(n-1)^-},\ldots,j_{1^-};j_{1^+},\ldots,j_{(n-1)^+},0)
    =-\frac{\ii\Delta t}{2}W^{(\tilde r)}
 Q_{n-1}^{(\tilde r)}(j_{(n-1)^-},\ldots,j_{1^-};j_{1^+},\ldots,j_{(n-1)^+}),\\
 &A_n^{(1/2)}(0,j_{(n-1)^-},\ldots,j_{1^-};j_{1^+},\ldots,j_{(n-1)^+},1)
    =\frac{\ii\Delta t}{2}
 Q_{n-1}^{(\tilde r)}(j_{(n-1)^-},\ldots,j_{1^-};j_{1^+},\ldots,j_{(n-1)^+})W^{(\tilde r)}.
 \end{aligned}
 \label{eq:generalhalfcoefficients}
\end{equation}
The second stage then updates $A_n^{(\tilde r)}$ with indices satisfying \eqref{eq:m_condition}:
\paragraph{Case 1: $\boldsymbol{j_{n^-} = j_{n^+} = 0}$}
\begin{equation}
 \begin{aligned}
 &A_n^{(\tilde r)}(0,j_{(n-1)^-},\ldots,j_{1^-};j_{1^+},\ldots,j_{(n-1)^+},0)\approx Q_{n-1}^{(\tilde r)}(j_{(n-1)^-},\ldots,j_{1^-};j_{1^+},\ldots,j_{(n-1)^+})\\
 &\qquad{}-\frac{\ii}{\Delta t}\sum_{k=1}^{n-\tilde r-1}\eta_{n-k}\Bigl[W^{(\tilde r)},A_n^{(1/2)}(0,\hat{j}_{(n-1)^-}^{k^-},\ldots,\hat{j}_{1^-}^{k^-};\hat{j}_{1^+}^{k^-},\ldots,\hat{j}_{(n-1)^+}^{k^-},0)\Bigr]\\
 &\qquad{}-\frac{\ii}{\Delta t}\sum_{k=1}^{n-\tilde r-1}\eta_{n-k}^*\Bigl[W^{(\tilde r)},A_n^{(1/2)}(0,\hat{j}_{(n-1)^-}^{k^+},\ldots,\hat{j}_{1^-}^{k^+};\hat{j}_{1^+}^{k^+},\ldots,\hat{j}_{(n-1)^+}^{k^+},0)\Bigr].
 \end{aligned}
 \label{eq:general-full-zero}
\end{equation}
\paragraph{Case 2: $\boldsymbol{j_{n^-}+j_{n^+}=1}$}
\begin{equation}
 \begin{aligned}
 &A_n^{(\tilde r)}(1,j_{(n-1)^-},\ldots,j_{1^-};j_{1^+},\ldots,j_{(n-1)^+},0)\approx -\ii\Delta t\,W^{(\tilde r)}A_n^{(1/2)}(0,j_{(n-1)^-},\ldots,j_{1^-};j_{1^+},\ldots,j_{(n-1)^+},0)\\
 &\qquad{}-\frac{\ii}{\Delta t}\sum_{k=1}^{n-\tilde r-1}\eta_{n-k}\Bigl[W^{(\tilde r)},A_n^{(1/2)}(1,\hat{j}_{(n-1)^-}^{k^-},\ldots,\hat{j}_{1^-}^{k^-};\hat{j}_{1^+}^{k^-},\ldots,\hat{j}_{(n-1)^+}^{k^-},0)\Bigr]\\
 &\qquad{}-\frac{\ii}{\Delta t}\sum_{k=1}^{n-\tilde r-1}\eta_{n-k}^*\Bigl[W^{(\tilde r)},A_n^{(1/2)}(1,\hat{j}_{(n-1)^-}^{k^+},\ldots,\hat{j}_{1^-}^{k^+};\hat{j}_{1^+}^{k^+},\ldots,\hat{j}_{(n-1)^+}^{k^+},0)\Bigr],\\
 &A_n^{(\tilde r)}(0,j_{(n-1)^-},\ldots,j_{1^-};j_{1^+},\ldots,j_{(n-1)^+},1) = [A_n^{(\tilde r)}(1,j_{(n-1)^+},\ldots,j_{1^+};j_{1^-},\ldots,j_{(n-1)^-},0)]^{\dagger}.
 \end{aligned}
 \label{eq:general-full-minus}
\end{equation}
\paragraph{Case 3: $\boldsymbol{j_{n^-}+j_{n^+}=2}$}
\begin{equation}
  \begin{aligned}
    A_n^{(\tilde r)}(2,j_{(n-1)^-},\ldots,j_{1^-};j_{1^+},\ldots,j_{(n-1)^+},0)&\approx -2\ii\Delta t\,W^{(\tilde r)}A_n^{(1/2)}(1,j_{(n-1)^-},\ldots,j_{1^-};j_{1^+},\ldots,j_{(n-1)^+},0),\\
    A_n^{(\tilde r)}(0,j_{(n-1)^-},\ldots,j_{1^-};j_{1^+},\ldots,j_{(n-1)^+},2)&=[A_n^{(\tilde r)}(2,j_{(n-1)^+},\ldots,j_{1^+};j_{1^-},\ldots,j_{(n-1)^-},0)]^{\dagger}, \\
    A_n^{(\tilde r)}(1,j_{(n-1)^-},\ldots,j_{1^-};j_{1^+},\ldots,j_{(n-1)^+},1)&\approx -\ii\Delta t\,W^{(\tilde r)}A_n^{(1/2)}(0,j_{(n-1)^-},\ldots,j_{1^-};j_{1^+},\ldots,j_{(n-1)^+},1)\\
    &\quad{}+\ii\Delta t\,A_n^{(1/2)}(1,j_{(n-1)^-},\ldots,j_{1^-};j_{1^+},\ldots,j_{(n-1)^+},0)W^{(\tilde r)}.
  \end{aligned}
 \label{eq:general-full-two}
\end{equation}
Equations \eqref{eq:generaltransfer}--\eqref{eq:general-full-two} define the iteration for $A_n^{(\tilde r)}$ and $Q_{n-1}^{(\tilde r)}$ in the second-order scheme. The initial condition is given by
\begin{equation}
  Q_0^{(0)}(\,;\,)=\mathcal E^{(0)}[\widetilde{\rho}_{s,0}],
  \label{eq:generalstart}
\end{equation}
and, once $A_n^{(\tilde r)}$ is computed, we have
\begin{equation}
  \widetilde{\rho}_s(n\Delta t)
  \approx\sum_{i_{n-\tilde r}^-,\ldots,i_n^-=1}^{M}
  \sum_{i_{n-\tilde r}^+,\ldots,i_n^+=1}^{M}
  \mel{i_n^-,\ldots,i_{n-\tilde r}^-}{A_n^{(\tilde r)}(0,\ldots,0;0,\ldots,0)}{i_n^+,\ldots,i_{n-\tilde r}^+}\dyad{i_n^-}{i_n^+}.
  \label{eq:generalreduceddensity}
\end{equation}
This approximation is expected to be second-order in $\Delta t$.

\subsection{Cost reduction for \texorpdfstring{$r\ge1$}{r>=1}}
\label{sec:costreduction}
The previous section provides a complete second-order scheme for any near-field radius $r \geqslant 0$, in which tensors of order $2(r+1)$ need to be stored during the computation. In this section, we show a simple trick to reduce the tensor order to $2r$ for $r \geqslant 1$, without any further approximation. 

For $n\ge r+1$, define
\begin{equation}
 \begin{aligned}
 & \overline A_n^{(r-1)}(j_{n^-},\ldots,j_{1^-};j_{1^+},\ldots,j_{n^+}) \\
 &\quad=\sum_{i_{n-r}^-,\ldots,i_n^-=1}^M
 \sum_{i_{n-r}^+,\ldots,i_n^+=1}^M
 \mel{i_n^-,\ldots,i_{n-r}^-}
 {A_n^{(r)}(j_{n^-},\ldots,j_{1^-};j_{1^+},\ldots,j_{n^+})}
 {i_n^+,\ldots,i_{n-r}^+} \\[-8pt]
 & \hspace{250pt} \dyad{i_n^-,\ldots,i_{n-r+1}^-}{i_n^+,\ldots,i_{n-r+1}^+}, \\
 & \overline Q_{n-1}^{(r-1)}(j_{(n-1)^-},\ldots,j_{1^-};j_{1^+},\ldots,j_{(n-1)^+}) \\
 &\quad=\sum_{i_{n-r}^-,\ldots,i_n^-=1}^M
 \sum_{i_{n-r}^+,\ldots,i_n^+=1}^M
 \mel{i_n^-,\ldots,i_{n-r}^-}
 {Q_{n-1}^{(r)}(j_{(n-1)^-},\ldots,j_{1^-};j_{1^+},\ldots,j_{(n-1)^+})}
 {i_n^+,\ldots,i_{n-r}^+} \\[-8pt]
 & \hspace{250pt} \dyad{i_n^-,\ldots,i_{n-r+1}^-}{i_n^+,\ldots,i_{n-r+1}^+}.
 \end{aligned}
 \label{eq:reducedcoefficient}
\end{equation}
Both are tensors of order $2r$. Combining \eqref{eq:generaltransfer} with
\eqref{eq:reducedcoefficient} gives, for $n\ge r+1$,
\begin{equation}
 \begin{aligned}
 &\overline Q_{n-1}^{(r-1)}(j_{(n-1)^-},\ldots,j_{1^-};j_{1^+},\ldots,j_{(n-1)^+})\\
 &\quad=\overline{\mathcal E}^{(r)}\!\Bigl[\widetilde V^{(r)}\,
 \overline A_{n-1}^{(r-1)}(j_{(n-1)^-},\ldots,j_{1^-};j_{1^+},\ldots,j_{(n-1)^+})
 (\widetilde V^{(r)})^{\dagger}\Bigr].
 \end{aligned}
 \label{eq:reducedtransferoperator}
\end{equation}
Here $\widetilde V^{(r)}$ is defined in \eqref{eq:historytransfermatrix}, with $n-1=r$, and the map $\overline{\mathcal E}^{(r)}$ is given by
\begin{equation}
 \begin{aligned}
 \overline{\mathcal E}^{(r)}[X]
 &=\sum_{i_0^-,\ldots,i_r^-=1}^M\sum_{i_0^+,\ldots,i_r^+=1}^M
 \exp\!\left[-(w_{i_0^-}-w_{i_0^+})
 \sum_{l=0}^{r}(\eta_l w_{i_l^-}-\eta_l^*w_{i_l^+})\right]\\
 &\quad{}\times\mel{i_0^-,\ldots,i_r^-}{X}{i_0^+,\ldots,i_r^+}
 \dyad{i_0^-,\ldots,i_{r-1}^-}{i_0^+,\ldots,i_{r-1}^+},
 \end{aligned}
 \label{eq:reducednearmap}
\end{equation}
which maps operators on $(\mathbb C^M)^{\otimes(r+1)}$ to operators on $(\mathbb C^M)^{\otimes r}$.

The iteration can be performed directly on these reduced tensors, without ever forming the larger tensors $A_n^{(r)}$ and $Q_{n-1}^{(r)}$. For the first $r$ steps, we use the original iteration to compute $A_r^{(r-1)}$, then initialize the reduced iteration by setting $\overline A_r^{(r-1)}=A_r^{(r-1)}$. For $n\ge r+1$, the algorithm continues as follows:
\begin{itemize}
  \item \textbf{Stage 1:} Compute $\overline Q_{n-1}^{(r-1)}$ by
  \eqref{eq:reducedtransferoperator}, and then form the intermediate coefficients:
\begin{equation}
 \begin{aligned}
 &\overline A_n^{(1/2)}(0,j_{(n-1)^-},\ldots,j_{1^-};j_{1^+},\ldots,j_{(n-1)^+},0)\\
 &\quad=\overline Q_{n-1}^{(r-1)}(j_{(n-1)^-},\ldots,j_{1^-};j_{1^+},\ldots,j_{(n-1)^+})\\
 &\qquad{}-\frac{\ii}{2\Delta t}\sum_{k=1}^{n-r-1}\eta_{n-k}
 \Bigl[W^{(r-1)},\overline Q_{n-1}^{(r-1)} (\hat j_{(n-1)^-}^{k^-},\ldots,\hat j_{1^-}^{k^-}; \hat j_{1^+}^{k^-},\ldots,\hat j_{(n-1)^+}^{k^-})\Bigr]\\
 &\qquad{}-\frac{\ii}{2\Delta t}\sum_{k=1}^{n-r-1}\eta_{n-k}^*
 \Bigl[W^{(r-1)},\overline Q_{n-1}^{(r-1)} (\hat j_{(n-1)^-}^{k^+},\ldots,\hat j_{1^-}^{k^+}; \hat j_{1^+}^{k^+},\ldots,\hat j_{(n-1)^+}^{k^+})\Bigr],\\
 &\overline A_n^{(1/2)}(1,j_{(n-1)^-},\ldots,j_{1^-};j_{1^+},\ldots,j_{(n-1)^+},0)\\
 &\quad=-\frac{\ii\Delta t}{2}W^{(r-1)} \overline Q_{n-1}^{(r-1)}(j_{(n-1)^-},\ldots,j_{1^-};j_{1^+},\ldots,j_{(n-1)^+}),\\
 &\overline A_n^{(1/2)}(0,j_{(n-1)^-},\ldots,j_{1^-};j_{1^+},\ldots,j_{(n-1)^+},1)\\
 &\quad=\frac{\ii\Delta t}{2} \overline Q_{n-1}^{(r-1)}(j_{(n-1)^-},\ldots,j_{1^-};j_{1^+},\ldots,j_{(n-1)^+})W^{(r-1)}.
 \end{aligned}
 \label{eq:reducedhalfcoefficients}
\end{equation}
  \item \textbf{Stage 2:} Compute coefficients $\overline A_n^{(r-1)}$ for the following three cases:
    \begin{itemize}
      \item \textbf{Case 1:} $\boldsymbol{j_{n^-}=j_{n^+}=0}$
\begin{equation}
 \begin{aligned}
 &\overline A_n^{(r-1)}(0,j_{(n-1)^-},\ldots,j_{1^-};j_{1^+},\ldots,j_{(n-1)^+},0)\\
 &\quad\approx \overline Q_{n-1}^{(r-1)}(j_{(n-1)^-},\ldots,j_{1^-};j_{1^+},\ldots,j_{(n-1)^+})\\
 &\qquad{}-\frac{\ii}{\Delta t}\sum_{k=1}^{n-r-1}\eta_{n-k}
 \Bigl[W^{(r-1)},\overline A_n^{(1/2)}(0,\hat{j}_{(n-1)^-}^{k^-},\ldots,\hat{j}_{1^-}^{k^-};\hat{j}_{1^+}^{k^-},\ldots,\hat{j}_{(n-1)^+}^{k^-},0)\Bigr]\\
 &\qquad{}-\frac{\ii}{\Delta t}\sum_{k=1}^{n-r-1}\eta_{n-k}^*
 \Bigl[W^{(r-1)},\overline A_n^{(1/2)}(0,\hat{j}_{(n-1)^-}^{k^+},\ldots,\hat{j}_{1^-}^{k^+};\hat{j}_{1^+}^{k^+},\ldots,\hat{j}_{(n-1)^+}^{k^+},0)\Bigr].
 \end{aligned}
 \label{eq:reduced-full-zero}
\end{equation}
      \item \textbf{Case 2:} $\boldsymbol{j_{n^-}+j_{n^+}=1}$
\begin{equation}
 \begin{aligned}
 &\overline A_n^{(r-1)}(1,j_{(n-1)^-},\ldots,j_{1^-};j_{1^+},\ldots,j_{(n-1)^+},0)\\
 &\quad\approx -\ii\Delta t\,W^{(r-1)}\overline A_n^{(1/2)}(0,j_{(n-1)^-},\ldots,j_{1^-};j_{1^+},\ldots,j_{(n-1)^+},0)\\
 &\qquad{}-\frac{\ii}{\Delta t}\sum_{k=1}^{n-r-1}\eta_{n-k}
 \Bigl[W^{(r-1)},\overline A_n^{(1/2)}(1,\hat{j}_{(n-1)^-}^{k^-},\ldots,\hat{j}_{1^-}^{k^-};\hat{j}_{1^+}^{k^-},\ldots,\hat{j}_{(n-1)^+}^{k^-},0)\Bigr]\\
 &\qquad{}-\frac{\ii}{\Delta t}\sum_{k=1}^{n-r-1}\eta_{n-k}^*
 \Bigl[W^{(r-1)},\overline A_n^{(1/2)}(1,\hat{j}_{(n-1)^-}^{k^+},\ldots,\hat{j}_{1^-}^{k^+};\hat{j}_{1^+}^{k^+},\ldots,\hat{j}_{(n-1)^+}^{k^+},0)\Bigr],\\
 &\overline A_n^{(r-1)}(0,j_{(n-1)^-},\ldots,j_{1^-};j_{1^+},\ldots,j_{(n-1)^+},1)\\
 &\quad= [\overline A_n^{(r-1)}(1,j_{(n-1)^+},\ldots,j_{1^+};j_{1^-},\ldots,j_{(n-1)^-},0)]^{\dagger}.
 \end{aligned}
 \label{eq:reduced-full-minus}
\end{equation}
      \item \textbf{Case 3:} $\boldsymbol{j_{n^-}+j_{n^+}=2}$
\begin{equation}
  \begin{aligned}
 &\overline A_n^{(r-1)}(2,j_{(n-1)^-},\ldots,j_{1^-};j_{1^+},\ldots,j_{(n-1)^+},0)\\
 &\quad\approx -2\ii\Delta t\,W^{(r-1)}\overline A_n^{(1/2)}(1,j_{(n-1)^-},\ldots,j_{1^-};j_{1^+},\ldots,j_{(n-1)^+},0),\\
 &\overline A_n^{(r-1)}(0,j_{(n-1)^-},\ldots,j_{1^-};j_{1^+},\ldots,j_{(n-1)^+},2)\\
 &\quad=[\overline A_n^{(r-1)}(2,j_{(n-1)^+},\ldots,j_{1^+};j_{1^-},\ldots,j_{(n-1)^-},0)]^{\dagger},\\
 &\overline A_n^{(r-1)}(1,j_{(n-1)^-},\ldots,j_{1^-};j_{1^+},\ldots,j_{(n-1)^+},1)\\
 &\quad\approx -\ii\Delta t\,W^{(r-1)}\overline A_n^{(1/2)}(0,j_{(n-1)^-},\ldots,j_{1^-};j_{1^+},\ldots,j_{(n-1)^+},1)\\
 &\quad\quad{}+\ii\Delta t\,\overline A_n^{(1/2)}(1,j_{(n-1)^-},\ldots,j_{1^-};j_{1^+},\ldots,j_{(n-1)^+},0)W^{(r-1)}.
 \end{aligned}
 \label{eq:reduced-full-two}
\end{equation}
    \end{itemize}
\end{itemize}
Both stages follow from the original iteration by summing over the oldest system pair of indices, as in \eqref{eq:reducedcoefficient}. Again, any quantity $\overline{A}_n^{(r-1)}$ with indices not satisfying \eqref{eq:m_condition} is set to zero.

For $n\ge r+1$, the reduced density matrix is obtained by
\begin{equation}
 \begin{aligned}
  \widetilde{\rho}_s(n\Delta t)
  &\approx\sum_{i_{n-r+1}^-,\ldots,i_n^-=1}^{M}
  \sum_{i_{n-r+1}^+,\ldots,i_n^+=1}^{M}\\
  &\quad{}\times\mel{i_n^-,\ldots,i_{n-r+1}^-}{\overline A_n^{(r-1)}(0,\ldots,0;0,\ldots,0)}{i_n^+,\ldots,i_{n-r+1}^+}\dyad{i_n^-}{i_n^+}.
 \end{aligned}
 \label{eq:reduceddensitycost}
\end{equation}

The cost reduction introduced above will be extremely useful especially when $M$ is large, which will reduce the storage cost by a factor of $M^2$.

\subsection{Complexity Analysis}
We now analyze the time and space complexity of the algorithm using the reduced operators introduced in Section~\ref{sec:costreduction} for $r\ge1$, together with the $r=0$ scheme in Section~\ref{sec:selfinteraction}.

Under \eqref{eq:m_condition}, there are $3^n$ index lists with degree at most one at every time step and $n\cdot3\cdot3^{n-1}$ lists with degree two at exactly one time step. Thus, for $r\ge1$ and $n\ge r+1$, there are $(n+1)3^n$ coefficients $\overline A_n^{(r-1)}(j_{n^-},\ldots,j_{1^-};j_{1^+},\ldots,j_{n^+})$. Each coefficient is an operator on $(\mathbb C^M)^{\otimes r}$ with $M^{2r}$ entries. Storing $\overline A_n^{(r-1)}$, $\overline Q_{n-1}^{(r-1)}$, and $\overline A_n^{(1/2)}$ therefore requires memory of size $O((n+1)3^nM^{2r})$. The first $r$ steps use operators with at most $M^{2r}$ entries, so for $N\ge r+1$ the peak coefficient storage over $N$ steps is
\begin{equation}
  O\!\left(N3^NM^{2r}\right),\qquad r\geqslant 1.
  \label{eq:generalspacecost}
\end{equation}
For $r=0$, each coefficient is an $M\times M$ matrix, giving the space complexity $O(N3^NM^2)$.

We now consider the time complexity. For $r\geqslant 1$ and $n\geqslant r+1$, the computational cost at each time step is listed in the following table:
\begin{center}
  \begin{tabular}{@{}lccc@{}}
    \toprule
    Operation & No. of values & Cost per value & Cost per step \\
    \midrule
    Compute $\overline Q_{n-1}^{(r-1)}$ \eqref{eq:reducedtransferoperator}
      & $n3^{n-1}$ & $O(M^{2(r+1)})$
      & $O(n3^{n-1}M^{2(r+1)})$ \\
    Compute $\overline A_n^{(1/2)}$, $\overline A_n^{(r-1)}$ \eqref{eq:reducedhalfcoefficients}--\eqref{eq:reduced-full-two}
      & $O(n3^n)$ & $O((n-r)M^{2r})$
      & $O(n(n-r)3^nM^{2r})$ \\
    Evaluate $\widetilde{\rho}_s(n\Delta t)$ \eqref{eq:reduceddensitycost}
      & $1$ & $O(M^{2r})$ & $O(M^{2r})$ \\
    \bottomrule
  \end{tabular}
\end{center}
For fixed $r\geqslant 1$ and $N\geqslant r+1$, summing these costs and including the first $r$ steps gives the total propagation work
\begin{equation}
  O\!\left(N3^NM^{2(r+1)}+N^23^NM^{2r}\right).
  \label{eq:generaltimecost}
\end{equation}
The two terms account for the computation of $\overline Q_{n-1}^{(r-1)}$ and the two stages computing $\overline A_n^{(r-1)}$, respectively.

For $r=0$, computing $Q_{n-1}^{(0)}$ by \eqref{eq:Q-coeff-r0} uses the dense matrix product $G_s^{\Delta t}A_{n-1}(G_s^{\Delta t})^\dagger$, costing $O(M^3)$ per coefficient, followed by the $O(M^2)$ application of $\mathcal{E}^{(0)}$. The two Taylor stages cost $O(nM^2)$ per coefficient. The total propagation work is therefore $O(N3^N(M^3+NM^2))$.

To conclude this section, we summarize the time and space complexities of the methods discussed above in Table~\ref{tab:complexity-comparison}. For fixed $r$, PIFFE has polynomial time and space complexity in the system dimension $M$, whereas the exponent of $M$ in QuAPI grows with the number of time steps $N$. Increasing $r$ retains more terms in exponential form but requires larger coefficient arrays and more work to compute $\overline Q_{n-1}^{(r-1)}$. Like FRODS, PIFFE with small $r$ can reduce memory use relative to QuAPI, particularly when $M$ is large. Even at $r=0$, PIFFE retains the influence terms with $j=k$ in exponential form and is therefore expected to improve accuracy over FRODS while maintaining comparable time complexity. Evidence will be provided in Section~\ref{sec:numerics} by numerical experiments.
\begin{table}[!htbp]
  \centering
  \caption{Space and time complexities over $N$ time steps for an $M$-level system.}
  \label{tab:complexity-comparison}
  \begin{tabular}{@{}lll@{}}
    \toprule
    Method & Space complexity & Time complexity \\
    \midrule
    QuAPI & $O(M^{2N})$ & $O(M^{2N})$ \\
    FRODS & $O(N3^NM^2)$ & $O(N 3^N (M^3+NM^2))$ \\
    PIFFE, $r=0$ & $O(N3^NM^2)$ & $O(N 3^N(M^3+NM^2))$ \\
    PIFFE, $r\ge1$ & $O(N3^NM^{2r})$ & $O(N3^NM^{2(r+1)}+N^23^NM^{2r})$ \\
    \bottomrule
  \end{tabular}
\end{table}

\section{Memory length truncation and limitation of the total degree}
\label{sec:truncation}
The number of values of $A_n^{(\tilde r)}$ stored in Section~\ref{sec:generalradius} grows with $n$. To reduce the computational and storage costs, we introduce a finite memory length and a bound on the sum of the indices $j_{k^-}$ and $j_{k^+}$. These two approximations are applied to the recurrence in \eqref{eq:generaltransfer}--\eqref{eq:general-full-two}.

\subsection{Memory length truncation}
\label{sec:memorytruncation}
As discussed in Section~\ref{sec:model}, the decay of the bath correlation function motivates neglecting $\eta_l$ for $l > \Kmax$. As a result, in the numerical scheme, the range of $k$ in the summations appearing in \eqref{eq:generalhalfcoefficients}--\eqref{eq:general-full-minus} (or \eqref{eq:reducedhalfcoefficients}--\eqref{eq:reduced-full-minus}) can be restricted:
\begin{equation}
  \sum_{k=1}^{n-\tilde r-1} \quad \longrightarrow \quad \sum_{k=\max(1,n-\Kmax)}^{n-\tilde r-1}.
  \label{eq:memorysums}
\end{equation}
Now, assume that we want to compute $A_N^{(\tilde r)}(0,\ldots,0;0,\ldots,0)$ with $N > \Kmax$, which can be obtained by setting $j_{1^{\pm}} = \cdots = j_{(N-1)^{\pm}} = 0$ in \eqref{eq:general-full-zero}. A close look at the right-hand side of \eqref{eq:general-full-zero} shows that $A_N^{(\tilde r)}(0,\ldots,0;0,\ldots,0)$ depends only on $A_{N-1}^{(\tilde r)}(j_{(N-1)^-},\ldots,j_{1^-};j_{1^+},\ldots,j_{(N-1)^+})$ with $j_{k^-}=j_{k^+}=0$ for all $1\le k < N-\Kmax$. Similarly, one can find that computing
\begin{displaymath}
  A_{N-1}^{(\tilde r)}(j_{(N-1)^-},\ldots,j_{(N-\Kmax)^-}, 0, \ldots, 0; 0, \ldots, 0, j_{(N-\Kmax)^+}, \ldots, j_{(N-1)^+})
\end{displaymath}
only requires $A_{N-2}^{(\tilde r)}(j_{(N-2)^-},\ldots,j_{1^-};j_{1^+},\ldots,j_{(N-2)^+})$ with $j_{k^-}=j_{k^+}=0$ for all $1\le k < N-\Kmax-1$. Continuing this argument recursively, we find that at any time step $n$, only $A_n^{(\tilde r)}(j_{n^-},\ldots,j_{1^-};j_{1^+},\ldots,j_{n^+})$ satisfying
\begin{equation}
  j_{k^-}=j_{k^+}=0,\qquad 1\le k\le n-\Kmax,
  \label{eq:memoryindices}
\end{equation}
is needed in future time steps, meaning that only $2\min(n,\Kmax)$ indices may be nonzero. Thus, each time step requires storing at most $(\Kmax+1) 3^{\Kmax}$ values of $A_n^{(\tilde r)}$, avoiding exponential growth of memory and time costs with respect to $n$. Similarly, when $r \geqslant 1$ and $n \geqslant r+1$, we only need to compute and store $\overline{A}_n^{(r-1)}(j_n^-,j_{n-1}^-, \ldots, j_1^-; j_1^+, \ldots, j_{n-1}^+, j_n^+)$ for all indices satisfying \eqref{eq:memoryindices}. The space complexity then becomes
\begin{equation}
  O\!\left(\Kmax 3^{\Kmax}M^{2\max(1,r)}\right)
\end{equation}
for all $0\leqslant r<\Kmax$.

When $n \leqslant \Kmax$, the replacement in \eqref{eq:memorysums} has no effect. For $n > \Kmax$, each sum in \eqref{eq:memorysums} contains only $\Kmax-r$ terms. Using the reduced operators from Section~\ref{sec:costreduction} for $r\ge1$, together with the $r=0$ scheme from Section~\ref{sec:selfinteraction}, the total propagation work over $N>\Kmax$ steps is
\begin{equation}
  \begin{cases}
    O\!\left(N\Kmax3^{\Kmax}\bigl[M^{2(r+1)}+(\Kmax-r+1)M^{2r}\bigr]\right),
      &1\le r<\Kmax,\\
    O\!\left(N\Kmax 3^{\Kmax}(M^3+\Kmax M^2)\right),
      &r=0<\Kmax.
  \end{cases}
\end{equation}
For $r\ge1$, the $M^{2(r+1)}$ term accounts for computing $\overline Q_{n-1}^{(r-1)}$ by \eqref{eq:reducedtransferoperator}. The $(\Kmax-r+1)M^{2r}$ term accounts for the two Taylor stages, including the $\Kmax-r$ far-field summands and the remaining operations per coefficient. The $M^3$ term for $r=0$ accounts for the dense matrix products used to compute $Q_{n-1}^{(0)}$. For fixed $\Kmax$, $r$, and $M$, the total time complexity grows only linearly in $N$.

One special case is $r = K_{\max} \geqslant 1$, where the summation in \eqref{eq:memorysums} becomes empty. Using the reduced operators from Section~\ref{sec:costreduction}, we obtain, for $n\ge r+1$,
\begin{equation}
  \begin{split}
    \overline A_n^{(r-1)}(0,\ldots,0;0,\ldots,0) &= \overline Q_{n-1}^{(r-1)}(0,\ldots,0;0,\ldots,0) \\
    &= \overline{\mathcal E}^{(r)}\!\Bigl[\widetilde V^{(r)}\overline A_{n-1}^{(r-1)}(0,\ldots,0;0,\ldots,0)(\widetilde V^{(r)})^\dagger\Bigr].
  \end{split}
\end{equation}
The reduced iteration again starts from $\overline A_r^{(r-1)}=A_r^{(r-1)}$, as in Section~\ref{sec:costreduction}. This matches exactly the iterative QuAPI method in \cite{makri1995}. In this case, we only need to consider quantities with all indices $j_{k^{\pm}} = 0$, so that the time and coefficient-storage complexities become $O(NM^{2(\Kmax+1)})$ and $O(M^{2\Kmax})$, respectively.

\subsection{Limitation of total degrees in the Taylor expansion}
\label{sec:degreebound}
Another way to reduce the computational cost, which has also been applied in \cite{sun2026}, is to truncate the series \eqref{eq:generalcoefficients} up to a given finite degree $D_{\max}$, which means only $A_n^{(\tilde r)}(j_{n^-},\ldots,j_{1^-};j_{1^+},\ldots,j_{n^+})$ or $\overline A_n^{(r-1)}(j_{n^-},\ldots,j_{1^-};j_{1^+},\ldots,j_{n^+})$ with
\begin{equation}
  \sum_{k=1}^{n}(j_{k^-}+j_{k^+})\le\Dmax
  \label{eq:totaldegreebound}
\end{equation}
will be computed and stored. All other values, in case they are encountered during the computation, will be regarded as zero. Note that to compute $A_n^{(\tilde r)}$ or $\overline{A}_n^{(r-1)}$ with $\sum_{k=1}^{n}(j_{k^-}+j_{k^+})=\Dmax$, we may need to use $A_n^{(1/2)}$ or $\overline{A}_n^{(1/2)}$ with $\sum_{k=1}^{n}(j_{k^-}+j_{k^+})=\Dmax+1$.

With this extra condition \eqref{eq:totaldegreebound} on the indices, the number of coefficients $A_n^{(\tilde r)}$ or $\overline{A}_n^{(r-1)}$ to be stored can be further reduced. In most of our computations, we will choose $K_{\max} + 1 \geqslant D_{\max}$, and here we will stick to this assumption. At time step $n$ with $n > K_{\max}$, the number of coefficients $A_n^{(\tilde{r})}$ or $\overline{A}_n^{(r-1)}$ to be computed and stored is
\begin{equation}
  N_{\mathrm{coef}}(K_{\max}, D_{\max}) = \sum_{d=0}^{\Dmax}2^d\binom{\Kmax}{d}
  +3\Kmax\sum_{d=0}^{\Dmax-2}2^d\binom{\Kmax-1}{d},
  \label{eq:coefficientcount}
\end{equation}
where the first sum counts coefficients with $j_{k^-}+j_{k^+}\le1$ at every time step. The second sum counts coefficients with $j_{k^-}+j_{k^+}=2$ at exactly one time step and at most one at all other steps. Without setting the maximum degree, the number of coefficients to be stored is $N_{\mathrm{coef}}(K_{\max}) = (K_{\max}+1)3^{K_{\max}}$.

\begin{figure}[!htbp]
  \centering
  \subfloat[$N_{\mathrm{coef}}(\Kmax,\Dmax)$\label{fig:coefficientcounts-absolute}]{\includegraphics[width=0.49\textwidth]{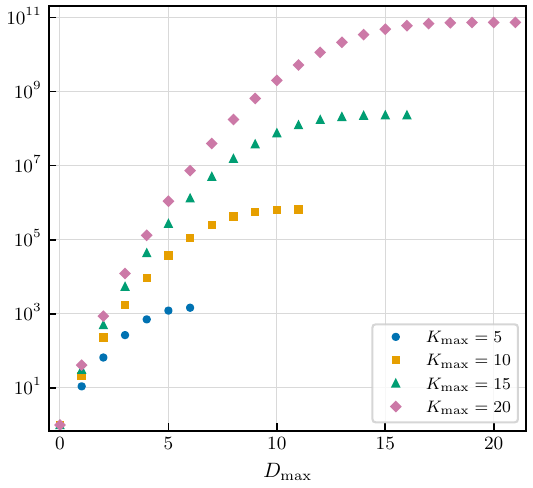}}
  \hfill
  \subfloat[$N_{\mathrm{coef}}(\Kmax,\Dmax)/N_{\mathrm{coef}}(\Kmax)$\label{fig:coefficientcounts-relative}]{\includegraphics[width=0.49\textwidth]{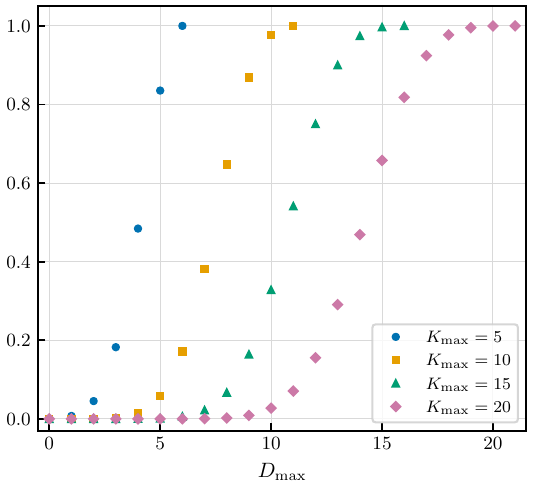}}
  \caption{Coefficient counts and their fractions of the counts without a degree cutoff, for $\Kmax=5,10,15,20$ and integer $0\le\Dmax\le\Kmax+1$. Panel (a) uses a logarithmic scale.}
  \label{fig:coefficientcounts}
\end{figure}

Figure~\ref{fig:coefficientcounts} compares these two counts for $\Kmax=5,10,15,20$. The left panel shows $N_{\mathrm{coef}}(\Kmax,\Dmax)$, which reaches $N_{\mathrm{coef}}(\Kmax)$ as $\Dmax$ increases; the right panel shows their ratio. Since \eqref{eq:m_condition} allows at most one time step with index sum two, the largest possible total degree is $\Kmax+1$. Smaller values of $\Dmax$ can greatly reduce storage: for example, at $\Kmax=20$ and $\Dmax=4$, the count decreases from $73\,222\,472\,421$ to $130\,821$, retaining approximately $1.79\times10^{-6}$ of the coefficients.

The bound \eqref{eq:totaldegreebound} truncates the multivariate Taylor series \eqref{eq:generalcoefficients} at total degree $\Dmax$. Higher-degree terms contribute to the reduced dynamics through higher powers of the interaction strength. When the system--bath interaction is weak, these contributions are small, so a low-degree Taylor approximation can capture the dominant effects with substantially fewer coefficients. In PIFFE, the near-field influence is retained in exponential form, so the effectiveness of this truncation depends on the strength of the remaining far-field interaction. Decay of the bath correlation can make this interaction weak even when the near-field interaction is appreciable. The choice of $\Dmax$ is checked by increasing it at fixed $\Delta t$, $\Kmax$, and $r$ and monitoring convergence of the reduced density matrix.

\section{Numerical results}
\label{sec:numerics}
In this section, we will validate PIFFE and demonstrate its efficiency through numerical experiments. In all our experiments, we consider an Ohmic bath with spectral density
\begin{equation}
 J(\omega)=\frac{\pi\xi}{2}\omega e^{-\omega/\omega_c},\qquad \omega>0,
 \label{eq:ohmicspectral}
\end{equation}
where $\xi$ controls the coupling strength and $\omega_c$ is the cutoff frequency. While the spectral density provides a continuous spectrum of bath modes, our experiments will use a discrete set of harmonic oscillators to represent the bath. The frequencies are
\begin{equation}
  \omega_j=-\omega_c\log\!\left[1-\frac{j}{L}
  (1-e^{-\omega_{\max}/\omega_c})\right],\qquad
  c_j=\omega_j\sqrt{\frac{\xi\omega_c}{L}
  (1-e^{-\omega_{\max}/\omega_c})},\quad j=1,\ldots,L,
  \label{eq:ohmic}
\end{equation}
where $\omega_{\max}$ is the maximum frequency and $L$ is the number of harmonic
oscillators. Thus, the corresponding thermal correlation function is
\begin{equation}
  B(t)=\frac12\sum_{j=1}^{L}\frac{c_j^2}{\omega_j}
  \left[\coth\!\left(\frac{\beta\omega_j}{2}\right)\cos(\omega_jt)
       -\ii\sin(\omega_jt)\right].
  \label{eq:ohmiccorrelation}
\end{equation}
We use $L=400$ throughout. Other parameters will be specified later.

In the numerical results to be shown below, only second-order methods are tested. It has been demonstrated in \cite{sun2026} that first-order methods are less accurate and stable. Three methods involved in the numerical experiments are
\begin{itemize}
  \item Iterative QuAPI (i-QuAPI): Parameters include the time step $\dt$ and the memory length $\Tmax=\Kmax\dt$.
  \item FRODS: Parameters include the time step $\dt$, the memory length $\Tmax=\Kmax\dt$, and the maximum degree $\Dmax$.
  \item PIFFE: Parameters include the time step $\dt$, the memory length $\Tmax=\Kmax\dt$, the maximum degree $\Dmax$, and the radius $r$ of the near-field interactions.
\end{itemize}

\subsection{Convergence for the spin--boson model}
\label{sec:spinconvergence}
In this section, we will monitor the convergence of PIFFE with respect to various parameters using the classical spin--boson model. The spin--boson model consists of a two-level system with
\begin{equation}
 H_s=\varepsilon\hat\sigma_z+\Delta\hat\sigma_x,\qquad
 W_s=\hat\sigma_z,\qquad
 \rho_s(0)=\dyad{\uparrow},
 \label{eq:spinmodel}
\end{equation}
where $\hat\sigma_z=\dyad{\uparrow} - \dyad{\downarrow}$ and $\hat\sigma_x=\dyad{\uparrow}{\downarrow} + \dyad{\downarrow}{\uparrow}$. We monitor the population difference $\langle\sigma_z(t)\rangle=\mel{\uparrow}{\rho_s(t)}{\uparrow} - \mel{\downarrow}{\rho_s(t)}{\downarrow}$, and set parameters as follows:
\begin{displaymath}
  \Delta = 1, \quad \omega_c = 5, \quad \omega_{\max} = 20, \quad \xi = 0.5, \quad \beta = 2.5, \quad T = 10, \quad \varepsilon = 0 \text{ or }1.
\end{displaymath}

\begin{figure}[!htbp]
 \centering
 \subfloat[Memory length $\Kmax$]{\includegraphics[width=.49\textwidth]{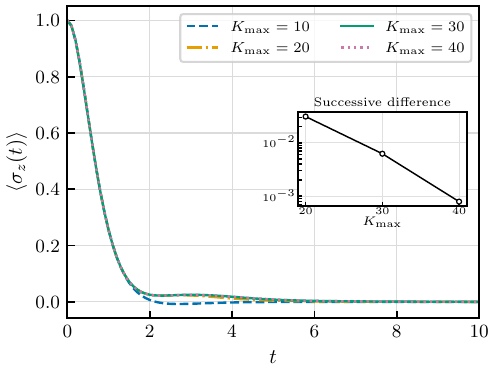}}
 \hfill
 \subfloat[Time step $\dt$]{\includegraphics[width=.49\textwidth]{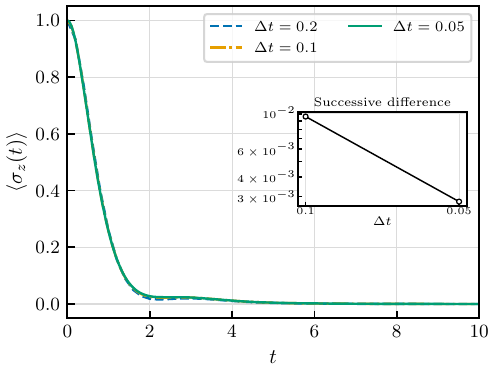}}\\
 \subfloat[Auxiliary degree $\Dmax$]{\includegraphics[width=.49\textwidth]{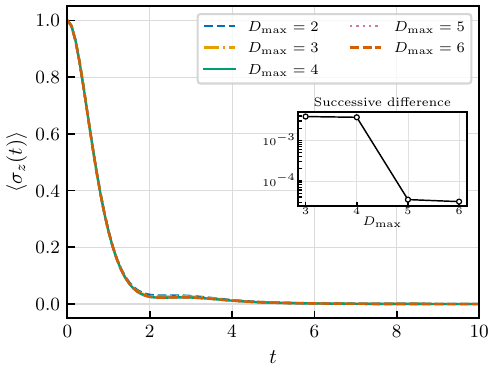}}
 \hfill
 \subfloat[Radius $r$]{\includegraphics[width=.49\textwidth]{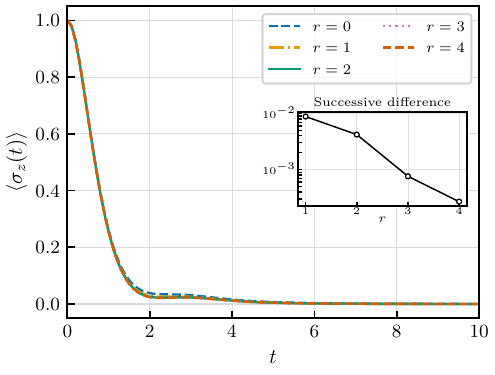}}
 \caption{PIFFE convergence tests for $\varepsilon=0$, $\Delta=1$,
 $\xi=0.5$, $\beta=2.5$, $\omega_c=5$, and $\omega_{\max}=20$. (a) $\dt=0.1$, $\Dmax=4$, $r=2$;
 (b) $\Tmax=2$, $\Dmax=4$, $r=2$;
 (c) $\dt=0.1$, $\Kmax=20$, $r=2$;
 (d) $\dt=0.1$, $\Kmax=20$, $\Dmax=4$.
 Insets show the maximum difference in $\langle\sigma_z(t)\rangle$ between
 successive refinements, plotted at the finer parameter value; time-step
 comparisons use common output times.}
 \label{fig:spinconvergence}
\end{figure}

\begin{figure}[!htbp]
 \centering
 \subfloat[Memory length $\Kmax$]{\includegraphics[width=.49\textwidth]{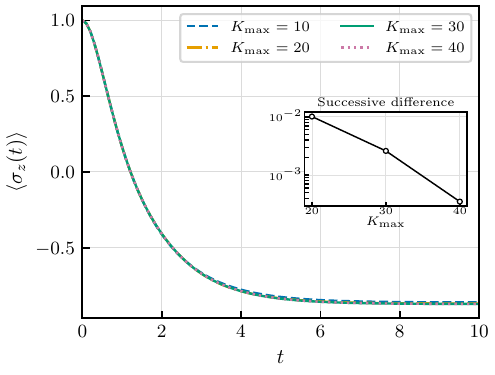}}
 \hfill
 \subfloat[Time step $\dt$]{\includegraphics[width=.49\textwidth]{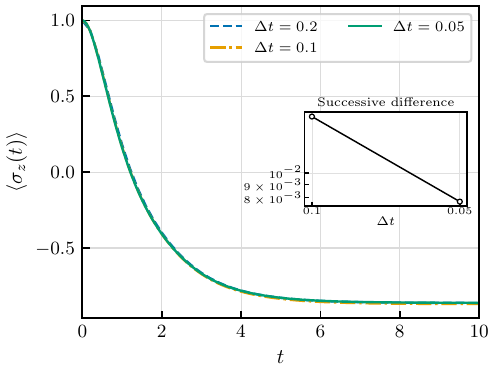}}\\
 \subfloat[Auxiliary degree $\Dmax$]{\includegraphics[width=.49\textwidth]{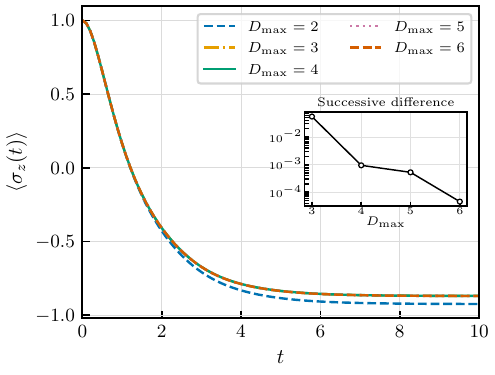}}
 \hfill
 \subfloat[Radius $r$]{\includegraphics[width=.49\textwidth]{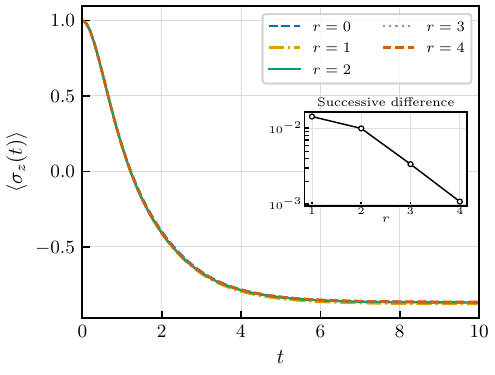}}
 \caption{PIFFE convergence tests for $\varepsilon=1$, $\Delta=1$,
 $\xi=0.5$, $\beta=2.5$, $\omega_c=5$, and $\omega_{\max}=20$. (a) $\dt=0.1$, $\Dmax=4$, $r=2$;
 (b) $\Tmax=2$, $\Dmax=4$, $r=2$;
 (c) $\dt=0.1$, $\Kmax=20$, $r=2$;
 (d) $\dt=0.1$, $\Kmax=20$, $\Dmax=4$.
 Insets show the maximum difference in $\langle\sigma_z(t)\rangle$ between
 successive refinements, plotted at the finer parameter value; time-step
 comparisons use common output times.}
 \label{fig:spinconvergencebiased}
\end{figure}

We examine the convergence with respect to the memory length $\Kmax$, the time step $\dt$, the maximum degree $\Dmax$, and the radius $r$. The results for $\varepsilon = 0$ and $\varepsilon = 1$ are given in Figures~\ref{fig:spinconvergence} and \ref{fig:spinconvergencebiased}, respectively. In each figure, the four panels show the convergence with respect to one of the four parameters. In detail,
\begin{itemize}
  \item Panel (a) shows the convergence with respect to the memory length $\Kmax=10,20,30,40$ at fixed $\dt=0.1$, $\Dmax=4$, and $r=2$.
  \item Panel (b) shows the convergence with respect to the time step $\dt$ using $\Kmax=10,20,40$ for $\dt=0.2,0.1,0.05$, respectively, so that the physical memory length remains fixed at $\Tmax=2$; $\Dmax=4$ and $r=2$.
  \item Panel (c) shows the convergence with respect to the maximum degree $\Dmax$ at fixed $\dt=0.1$, $\Kmax=20$, and $r=2$.
  \item Panel (d) shows the convergence with respect to the radius $r$ at fixed $\dt=0.1$, $\Kmax=20$, and $\Dmax=4$.
\end{itemize}
For each pair of adjacent refinement levels, the \emph{successive difference} in the insets means the maximum absolute difference in $\langle\sigma_z(t)\rangle$ over output times shared by both runs, with all other parameters fixed as specified above. For example, the point at $\Dmax=4$ compares the runs with $\Dmax=3$ and $4$; the point at $\dt=0.05$ compares the runs with $\dt=0.1$ and $0.05$.

The successive differences decrease under refinements of all the four parameters for both biases, showing the trend of convergence. Despite the absence of an exact solution, the converged solutions in all the four panels are consistent with each other, indicating that the numerical results are reliable. Thermal equilibrium is reached as $t\to\infty$. When $\varepsilon = 0$, symmetry gives equal spin populations, and hence $\langle\sigma_z(t)\rangle\to0$. For $\varepsilon=1$, the bias favors $\ket{\downarrow}$, so a negative limiting value is expected.

For a separate temporal-order check, we fix $r=1$, $\Dmax=4$, and use the full memory length. Let $\langle\sigma_z(T)\rangle_h$ denote the computed observable with step size $h$. We estimate the order by
\begin{equation}
 p=\log_2\frac{\bigl|\langle\sigma_z(T)\rangle_{4h}-\langle\sigma_z(T)\rangle_{2h}\bigr|}
 {\bigl|\langle\sigma_z(T)\rangle_{2h}-\langle\sigma_z(T)\rangle_h\bigr|},
 \qquad h=0.0125.
 \label{eq:observedorder}
\end{equation}
Table~\ref{tab:spinorder} gives the $p$ values at $T = 0.4$ for various values of the coupling intensity $\xi$ and inverse temperature $\beta$. All the values are close to $2$, agreeing with the expected second-order convergence, while larger deviations occur at stronger coupling between the system and the bath, requiring smaller time steps to achieve the expected numerical order.

\begin{table}[!htbp]
 \centering\small
 \setlength{\tabcolsep}{5pt}
 \caption{Temporal orders of PIFFE for the spin--boson model at $T=0.4$.}
 \label{tab:spinorder}
 \begin{tabular}{@{}rrrrr@{\qquad}rrrr@{}}\toprule
 & \multicolumn{4}{c}{$\varepsilon=0$} & \multicolumn{4}{c}{$\varepsilon=1$}\\
 \cmidrule(lr){2-5}\cmidrule(lr){6-9}
 $\beta$ & $\xi=0.25$ & $\xi=0.5$ & $\xi=0.75$ & $\xi=1$
 & $\xi=0.25$ & $\xi=0.5$ & $\xi=0.75$ & $\xi=1$\\\midrule
 $5$ & $1.9923$ & $1.9686$ & $1.9193$ & $1.8462$ & $1.9818$ & $1.9307$ & $1.8422$ & $1.7239$\\
 $2$ & $1.9958$ & $1.9813$ & $1.9442$ & $1.8821$ & $1.9837$ & $1.9379$ & $1.8576$ & $1.7486$\\
 $1$ & $2.0065$ & $2.0204$ & $2.0204$ & $1.9920$ & $1.9897$ & $1.9611$ & $1.9066$ & $1.8265$\\
 \bottomrule\end{tabular}
\end{table}

Tables~\ref{tab:peakmemoryK} and~\ref{tab:peakmemoryD} record the peak resident memory during the simulation. For the same $K_{\max}$, FRODS and PIFFE require much less memory compared to i-QuAPI, making the simulations feasible for longer memory lengths. Note that the measured memory usage of PIFFE does not increase as $O(M^{2\max(1,r)})$ as stated in Section~\ref{sec:memorytruncation}, due to the overhead storage to build the dependency relationship in the iterative scheme. Note that due to the reduced operators introduced in Section~\ref{sec:costreduction}, the storage requirement of PIFFE for $r = 1$ is almost the same as that for $r = 0$.

\begin{table}[!htbp]
 \centering\small
 \caption{Peak resident memory in MiB at $\Dmax=4$.}
 \label{tab:peakmemoryK}
 \begin{tabular}{@{}rrrrrr@{}}\toprule
 & & & \multicolumn{3}{c}{PIFFE}\\\cmidrule(l){4-6}
 $\Kmax$ & i-QuAPI & FRODS & $r=0$ & $r=1$ & $r=2$\\\midrule
 $8$ & $63.4$ & $7.2$ & $7.1$ & $7.2$ & $10.3$\\
 $10$ & $138.4$ & $10.4$ & $10.7$ & $9.9$ & $16.9$\\
 $12$ & $1338.7$ & $16.6$ & $16.4$ & $16.0$ & $29.8$\\
 $15$ & --- & $32.8$ & $32.7$ & $31.9$ & $63.6$\\
 $20$ & --- & $90.2$ & $90.3$ & $88.1$ & $184.2$\\
 \bottomrule\end{tabular}
\end{table}

\begin{table}[!htbp]
 \centering\small
 \caption{Peak resident memory in MiB at $\Kmax=12$.}
 \label{tab:peakmemoryD}
 \begin{tabular}{@{}rrrrrr@{}}\toprule
 & & & \multicolumn{3}{c}{PIFFE}\\\cmidrule(l){4-6}
 $\Dmax$ & i-QuAPI & FRODS & $r=0$ & $r=1$ & $r=2$\\\midrule
 $2$ & $1338.7$ & $4.6$ & $4.6$ & $4.6$ & $5.0$\\
 $3$ & $1338.7$ & $6.1$ & $6.2$ & $6.4$ & $8.4$\\
 $4$ & $1338.7$ & $16.6$ & $16.4$ & $16.0$ & $29.8$\\
 $5$ & $1338.7$ & $67.4$ & $67.3$ & $64.8$ & $133.5$\\
 \bottomrule\end{tabular}
\end{table}

\subsection{Dependence on the coupling strength}
\label{sec:couplingtest}
\begin{figure}[!htbp]
 \centering
 \subfloat[$\xi=0.25$]{\includegraphics[width=.49\textwidth]{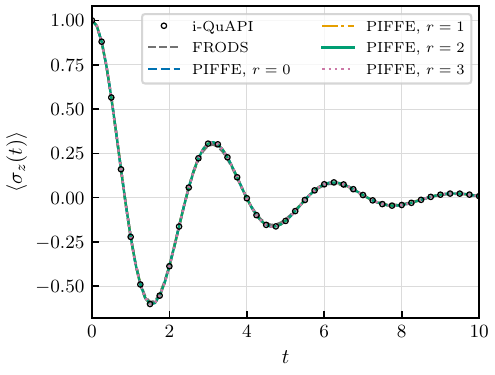}}
 \hfill
 \subfloat[$\xi=0.5$]{\includegraphics[width=.49\textwidth]{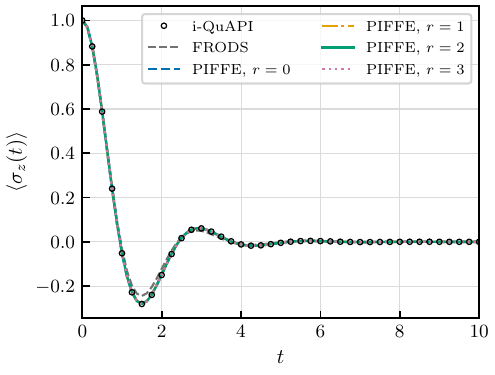}}\\
 \subfloat[$\xi=0.75$]{\includegraphics[width=.49\textwidth]{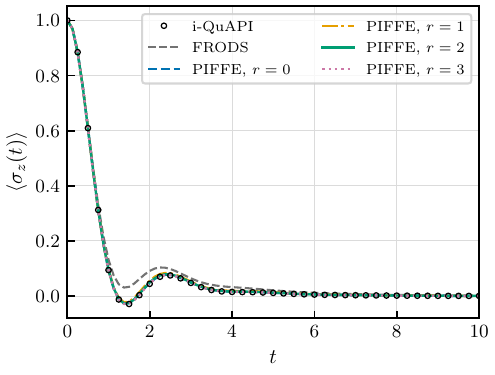}}
 \hfill
 \subfloat[$\xi=1$]{\includegraphics[width=.49\textwidth]{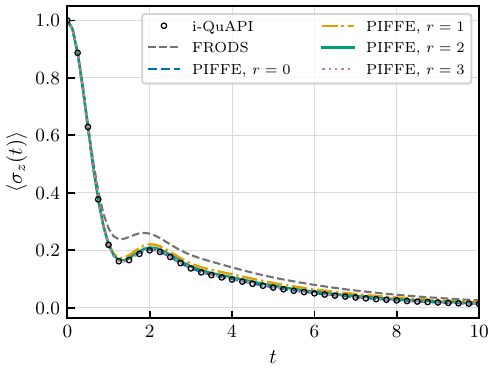}}
 \caption{Coupling-strength comparison at $\beta=5$, $\dt=0.125$,
 $\Kmax=12$, and $\Dmax=4$.}
 \label{fig:couplings}
\end{figure}

We return to $\varepsilon=0$ and take $\xi=0.25,0.5,0.75,1$, with $\beta=5$ and $T=10$, and study the performance of different methods (i-QuAPI, FRODS, and PIFFE) at various coupling strengths. The bath parameters are $\omega_c=2.5$ and $\omega_{\max}=10$. For all the methods, we use the same time step $\dt=0.125$ and the memory length $\Tmax=1.5$ (i.e., $\Kmax=12$). Both FRODS and PIFFE use $\Dmax=4$. The evolution of $\langle\sigma_z(t)\rangle$ is shown in Figure~\ref{fig:couplings}.

Figure \ref{fig:couplings} shows that the FRODS results gradually lose accuracy as the coupling strength increases, while the i-QuAPI results and PIFFE results remain consistent with each other, even for the radius $r = 0$. Nevertheless, the PIFFE runs require much less memory compared to i-QuAPI, as shown in Table \ref{tab:peakmemorycoupling}.

\begin{table}[!htbp]
  \centering\small
  \caption{Peak resident memory in MiB for the coupling-strength comparison
  in Figure~\ref{fig:couplings}. Each simulation is measured separately:
  QuAPI and FRODS use one CPU worker, and PIFFE uses two.
  The QuAPI memory measurement at $\xi=0.5$ is repeated in all four rows,
  since its array dimensions are independent of $\xi$.}
  \label{tab:peakmemorycoupling}
  \begin{tabular}{@{}rrrrrrr@{}}\toprule
  & & & \multicolumn{4}{c}{PIFFE}\\\cmidrule(l){4-7}
  $\xi$ & QuAPI & FRODS & $r=0$ & $r=1$ & $r=2$ & $r=3$\\\midrule
  $0.25$ & $1338.7$ & $16.4$ & $16.4$ & $16.1$ & $29.9$ & $86.1$\\
  $0.5$ & $1338.7$ & $16.3$ & $16.6$ & $16.1$ & $29.8$ & $86.6$\\
  $0.75$ & $1338.7$ & $16.4$ & $16.4$ & $16.0$ & $29.6$ & $86.1$\\
  $1$ & $1338.7$ & $16.4$ & $16.3$ & $15.8$ & $29.8$ & $86.5$\\
  \bottomrule\end{tabular}
\end{table}

\begin{figure}[!htbp]
 \centering\includegraphics[width=.65\textwidth]{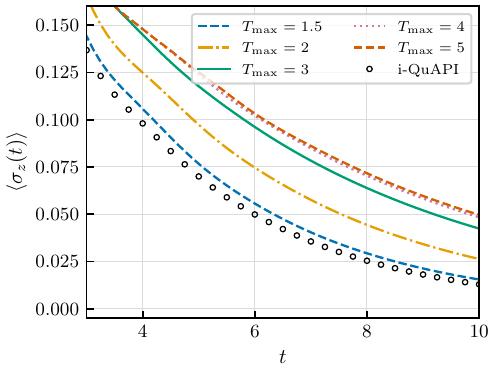}
 \caption{Memory extension with PIFFE for $\xi=1$, $\beta=5$, $r=2$,
 $\dt=0.125$, and $\Dmax=4$. The post-transient region is shown.
 Circles: i-QuAPI at the same step size with $\Tmax=1.5$.}
 \label{fig:extendedmemory}
\end{figure}

At $\xi = 1$, the results shown in Figure~\ref{fig:couplings} might be inaccurate due to the truncated memory length $\Tmax=1.5$. To check this, we increase the memory length to $\Tmax=2,3,4,5$ while keeping $\dt=0.125$, $r=2$, and $\Dmax=4$ fixed. At $\Tmax=2$ ($\Kmax=16$), the i-QuAPI implementation requires an estimated more than $300$ GiB, exceeding the available physical memory. Therefore, we only show PIFFE results at these longer memory lengths in Figure~\ref{fig:extendedmemory}. The calculation with the longest memory retains $\Kmax=40$ time steps and uses 2.61 GiB of peak resident memory on two CPU workers. The i-QuAPI curve in this figure retains $\Kmax=12$.

Figure~\ref{fig:extendedmemory} shows that increasing the memory length raises the value of $\langle\sigma_z(t)\rangle$ for $t \in [4, 10]$, indicating possible inaccuracy of the result in Figure~\ref{fig:couplings}(d). The curves for $\Tmax=4$ and $5$ are nearly indistinguishable, suggesting the convergence with respect to the memory length. This example shows that PIFFE makes simulations with longer memory more accessible with modest storage.

\subsection{Dependence on the inverse temperature}
\label{sec:temperaturetest}
We now fix $\xi=0.75$, $\Delta=1$, $\omega_c=2.5$, $\omega_{\max}=10$ and $\varepsilon=1$ and vary the inverse temperature over $\beta=5,2,1,0.5$. Decreasing $\beta$ increases the factor $\coth(\beta\omega/2)$ in~\eqref{eq:ohmiccorrelation}, which increases the value of $B(0)$, causing stronger thermal fluctuations.

\begin{figure}[!htbp]
 \centering
 \subfloat[$\beta=5$]{\includegraphics[width=.49\textwidth]{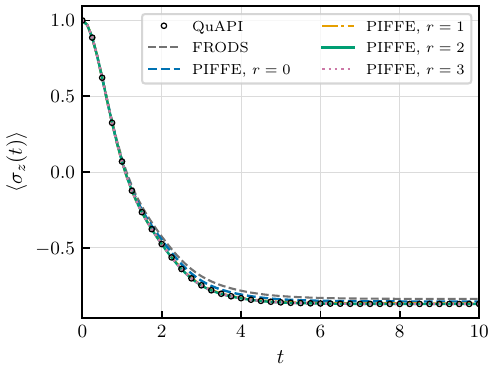}}
 \hfill
 \subfloat[$\beta=2$]{\includegraphics[width=.49\textwidth]{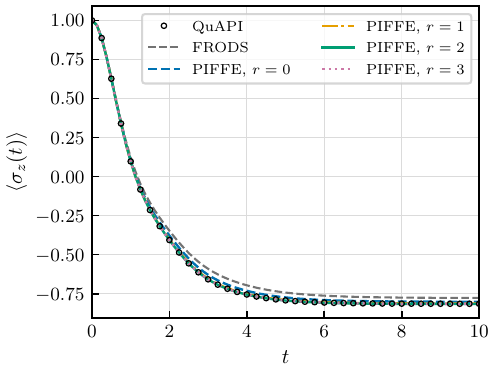}}\\
 \subfloat[$\beta=1$]{\includegraphics[width=.49\textwidth]{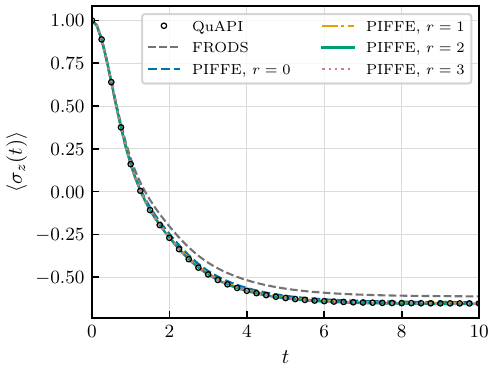}}
 \hfill
 \subfloat[$\beta=0.5$]{\includegraphics[width=.49\textwidth]{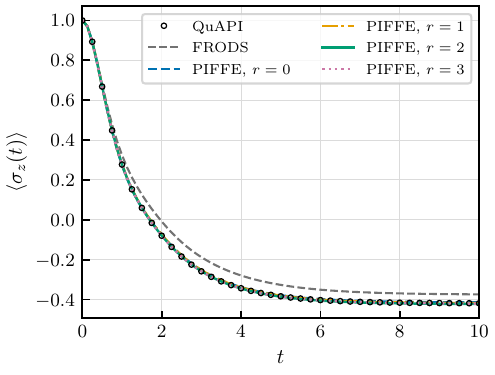}}
 \caption{Temperature comparison at $\xi=0.75$, $\varepsilon=1$,
 $\dt=0.125$, $\Kmax=12$, and $\Dmax=4$, with
 $\omega_c=2.5$, $\omega_{\max}=10$, and $\Delta=1$.}
 \label{fig:temperatures}
\end{figure}

In the numerical tests, we use $\dt=0.125$, $\Kmax=12$, $\Dmax=4$, and $T=10$.  Figure~\ref{fig:temperatures} compares FRODS, PIFFE, and QuAPI at each temperature. It is seen that for larger temperatures, the final population difference is closer to zero due to stronger thermal fluctuations. The general behaviors of FRODS and PIFFE are the same as in Section~\ref{sec:couplingtest}: FRODS deviates from QuAPI as the temperature increases, while PIFFE remains consistent with QuAPI even at $r=0$.

Quantitatively, we provide the maximum absolute differences between the QuAPI results and FRODS/PIFFE results in Table~\ref{tab:temperatureerrors}, defined by
\begin{equation}
 E_{\rm num}=\max_{0\le n\le N}
 \left|\langle\sigma_z(n\dt)\rangle_{\rm num}
       -\langle\sigma_z(n\dt)\rangle_{\rm QuAPI}\right|,
 \qquad N\dt=T,
 \label{eq:quapidifference}
\end{equation}
where $\langle\sigma_z(n\dt)\rangle_{\rm num}$ is the result from FRODS or PIFFE, and $\langle\sigma_z(n\dt)\rangle_{\rm QuAPI}$ is the result from QuAPI. All methods use the same time step and physical memory length $\Tmax=1.5$. For every row in Table~\ref{tab:temperatureerrors}, FRODS gives the largest difference from QuAPI, while the PIFFE error decreases as $r$ increases, except for $\beta=0.5$, where the $r=1$ error is slightly larger than the $r=0$ error, which is most likely coincidental.

\begin{table}[!htbp]
 \centering\small
 \caption{Maximum differences $E_{\rm num}$ defined in \eqref{eq:quapidifference} for Figure~\ref{fig:temperatures}.}
 \label{tab:temperatureerrors}
 \begin{tabular}{@{}rrrrrr@{}}\toprule
 & & \multicolumn{4}{c}{PIFFE}\\\cmidrule(l){3-6}
 $\beta$ & FRODS & $r=0$ & $r=1$ & $r=2$ & $r=3$\\\midrule
 $5$ & $0.0607$ & $0.032$ & $0.00328$ & $0.00109$ & $0.00031$\\
 $2$ & $0.0674$ & $0.0308$ & $0.0046$ & $0.00136$ & $0.000533$\\
 $1$ & $0.0693$ & $0.0227$ & $0.00664$ & $0.00306$ & $0.00144$\\
 $0.5$ & $0.0723$ & $0.00578$ & $0.00622$ & $0.00363$ & $0.00168$\\
 \bottomrule\end{tabular}
\end{table}

\subsection{Multilevel systems}
\label{sec:multileveltest}
Following Section 7.3 of~\cite{sun2026}, we consider an $M$-level system with nearest-neighbor coupling:
\begin{equation}
 H_s=\Omega\sum_{j=1}^{M-1}
 \bigl(|j\rangle\langle j+1|+|j+1\rangle\langle j|\bigr),\qquad
 W_s=\operatorname{diag}\left(-1,-1+\frac1J,\ldots,1\right),\quad
 J=\frac{M-1}{2}.
 \label{eq:chainmodel}
\end{equation}
We first consider $M = 11$ with
\begin{displaymath}
  \rho_s(0) = \dyad{1}, \quad \Omega = 1, \quad \beta = 5, \quad \omega_c = 2.5, \quad \xi = 0.4 \text{ and } 1.
\end{displaymath}
Using parameters $r = 0,1$ and $\Dmax=4$, we monitor all site populations $P_j(t)=\langle j|\rho_s(t)|j\rangle$ until $T=10$. The initial memory choice is $\Tmax=1$, with $\dt=0.05$ and $\Kmax=20$. Longer memory will be tested separately below.

\begin{figure}[!htbp]
 \centering
 \subfloat[$\xi=0.4$]{\includegraphics[width=.49\textwidth]{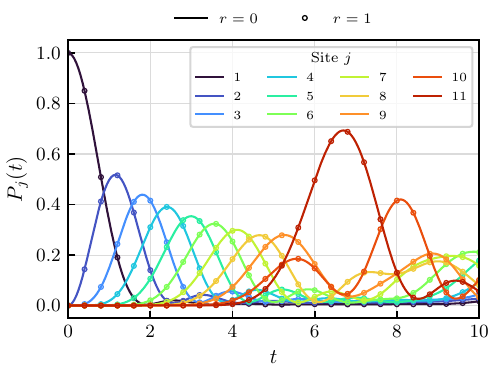}}
 \hfill
 \subfloat[$\xi=1$]{\includegraphics[width=.49\textwidth]{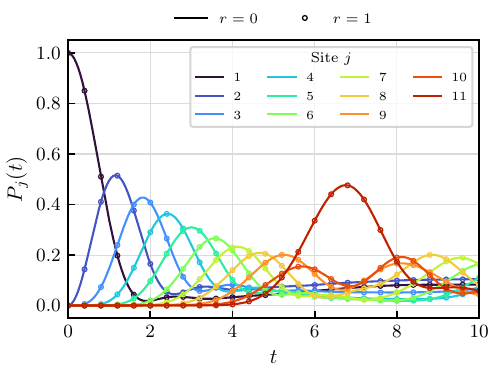}}
 \caption{PIFFE site populations for the eleven-level system, with $\beta=5$,
 $\Dmax=4$, $\Tmax=1$, $\dt=0.05$, and $\Kmax=20$.
 Lines show $r=0$ and circles show $r=1$. Colors identify the sites.}
 \label{fig:multilevel}
\end{figure}

The evolutions of the populations are plotted in Figure~\ref{fig:multilevel}. The $r=0$ and $r=1$ curves are visually indistinguishable: the largest differences between $r=0$ and $r=1$ are $1.14\times10^{-5}$ for $\xi=0.4$ and $7.18\times10^{-5}$ for $\xi=1$. Comparing the two coupling strengths, we observe lower peak populations for larger levels, indicating reduced transfer of population due to the system--bath interaction.

\begin{figure}[!htbp]
 \centering
 \subfloat[$r=0$]{\includegraphics[width=.49\textwidth]{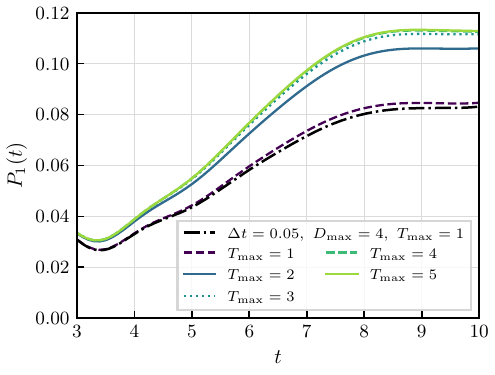}}
 \hfill
 \subfloat[$r=1$]{\includegraphics[width=.49\textwidth]{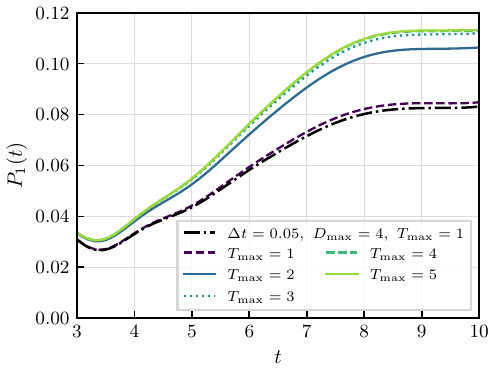}}
 \caption{Evolution of the first-level population for the 11-level system. The black dash-dotted curves use $\dt=0.05$, $\Dmax=4$, and $\Tmax=1$. Other curves use $\dt=0.1$ and $\Dmax=3$.}
 \label{fig:multilevelmemory}
\end{figure}

Inspired by the memory convergence tests in Section~\ref{sec:couplingtest}, we expect some improvement of accuracy by increasing the memory length in the case of $\xi=1$. The experiments are carried out for $\Tmax=1,2,3,4,5$ with a slightly larger time step $\dt = 0.1$ and a lower degree $\Dmax = 3$. For better visualization, we focus only on the population of the first level $P_1(t)$. One can see from Figure~\ref{fig:multilevelmemory} that using looser parameters $\dt=0.1$ and $\Dmax=3$ only slightly changes the population curves, but the memory length $\Tmax$ has a more significant effect. The curves show that $\Tmax$ needs to be at least $3$ to achieve converged dynamics of $P_1(t)$ up to $T=10$. 

The maximum population changes for all the 11 levels when increasing the memory length are recorded in Table~\ref{tab:multilevelmemory}. For the largest case with $T_{\max}=5$ ($K_{\max} = 50$), $r = 1$ and $D_{\max} = 3$, the peak resident memory is 1.35 GiB, and the total computational time is 248.38 s. Both the memory and time are feasible for a personal computer.

\begin{table}[!htbp]
 \centering\small
 \caption{Population changes for all eleven levels when increasing the
 memory length, with $\xi=1$, $\beta=5$, $r=0$, $\Dmax=3$, and $\dt=0.1$.
 Each entry is the maximum absolute change in $P_j(t)$ over common output
 times in $0\le t\le10$ for the indicated change in $\Tmax$.}
 \label{tab:multilevelmemory}
 \begin{tabular}{@{}rrrrr@{}}\toprule
 & \multicolumn{4}{c}{Memory change $\Tmax$}\\\cmidrule(l){2-5}
 Level $j$ & $1\to2$ & $2\to3$ & $3\to4$ & $4\to5$\\\midrule
 $1$ & $0.0215$ & $0.00572$ & $0.00141$ & $0.000292$\\
 $2$ & $0.0218$ & $0.00465$ & $0.000682$ & $0.000123$\\
 $3$ & $0.00521$ & $0.0023$ & $0.000745$ & $0.000232$\\
 $4$ & $0.0139$ & $0.00448$ & $0.000781$ & $0.000187$\\
 $5$ & $0.0199$ & $0.00583$ & $0.000785$ & $0.000138$\\
 $6$ & $0.0143$ & $0.00414$ & $0.00068$ & $0.000123$\\
 $7$ & $0.00955$ & $0.00194$ & $0.000547$ & $0.000167$\\
 $8$ & $0.0111$ & $0.00325$ & $0.000819$ & $0.000296$\\
 $9$ & $0.0136$ & $0.0045$ & $0.0015$ & $0.000512$\\
 $10$ & $0.0201$ & $0.00512$ & $0.00102$ & $0.000307$\\
 $11$ & $0.0142$ & $0.00266$ & $0.000325$ & $0.000105$\\
 \bottomrule\end{tabular}
\end{table}

\begin{figure}[!htbp]
 \centering
 \subfloat[No bath ($\xi=0$)]{\includegraphics[width=.49\textwidth]{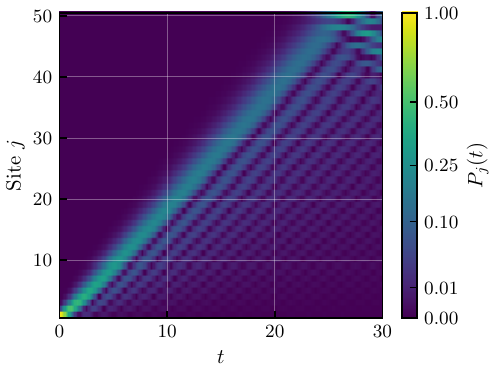}}
 \par\medskip
 \subfloat[$\Tmax=3$, $\Dmax=3$]{\includegraphics[width=.49\textwidth]{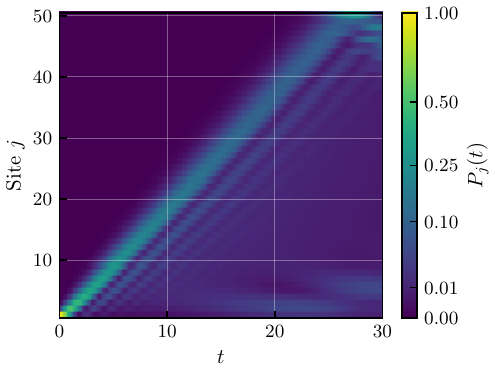}}
 \hfill
 \subfloat[$\Tmax=4$, $\Dmax=4$]{\includegraphics[width=.49\textwidth]{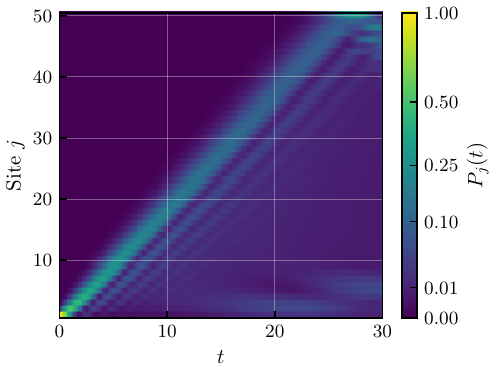}}
 \caption{Site populations for the fifty-level system with $\Omega=1$
 and $\rho_s(0)=\dyad{1}$, up to $T=30$.
 (a) Exact unitary evolution without the bath, sampled at intervals of $0.2$.
 (b,c) PIFFE with $\xi=1$, $\beta=5$, $r=1$, and $\dt=0.2$, for the
 indicated memory and degree cutoffs. All three panels use the same axes
 and color scale, proportional to $\sqrt{P_j}$, with labels showing $P_j$.}
 \label{fig:fiftylevel}
\end{figure}

In our last test, we set $M=50$ in~\eqref{eq:chainmodel}, retaining $\Omega=1$ and the initial state $\rho_s(0)=\dyad{1}$. The PIFFE simulations use $\dt=0.2$ and $r=1$, with two sets of memory and degree cutoffs:
\begin{itemize}
  \item $T_{\max} = 3$ ($K_{\max} = 15$) and $D_{\max} = 3$. The simulation uses eight OpenMP workers and takes $1919$ seconds ($32.0$ minutes) of total wall time. The peak resident memory is $1.062$ GiB.
  \item $T_{\max} = 4$ ($K_{\max} = 20$) and $D_{\max} = 4$. The simulation uses sixteen OpenMP workers and takes $37061$ seconds ($10.3$ hours) of total wall time. The peak resident memory is $20.238$ GiB.
\end{itemize}
Figure~\ref{fig:fiftylevel} shows site populations produced by the two simulations, where the results of an isolated system without bath are also provided for the purpose of comparison. The two PIFFE simulations are visually indistinguishable: increasing the memory and degree cutoffs together changes any site population by at most $1.71\times10^{-3}$ over $0\le t\le30$.

\begin{figure}[!htbp]
 \centering
 \includegraphics[width=.75\textwidth]{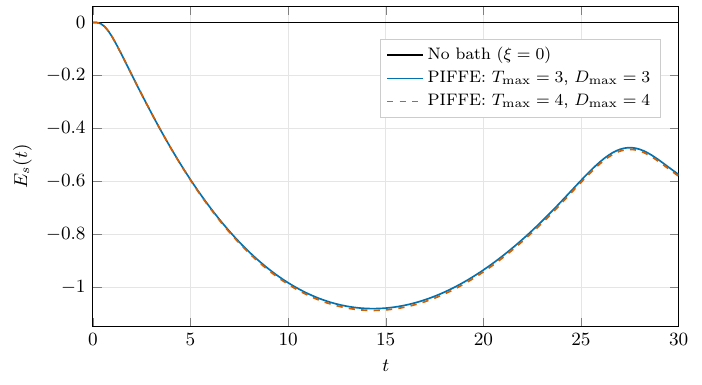}
 \caption{System energy $E_s(t)=\operatorname{Tr}[H_s\rho_s(t)]$}
 \label{fig:fiftylevelenergy}
\end{figure}

In all three panels, the initially localized population develops a moving front. The bath-coupled cases exhibit lower contrast in the oscillations behind the front, consistent with bath-induced changes in the intersite coherences. The front advances at approximately the same rate as in the isolated case, but more population remains at lower-index sites, reducing the mean distance traveled from the initial site. Figure~\ref{fig:fiftylevelenergy} shows the system energy $E_s(t)=\operatorname{Tr}[H_s\rho_s(t)]$ for the three simulations. The isolated system conserves $E_s(t)=0$, while both bath-coupled simulations show an initial energy decrease followed by partial recovery, indicating the effect of quantum dissipation.

\section{Conclusions}
\label{sec:conclusions}
We have established the relationship between QuAPI and FRODS by showing how FRODS-type iterative schemes can be derived from QuAPI through a Taylor expansion of the bath influence. Building on this connection, we have developed PIFFE by combining the exponential treatment of bath influence in QuAPI with the coefficient expansion of FRODS. Interactions spanning at most $r$ time steps are retained in exponential form, while the remaining far-field contribution is expanded to obtain iterative schemes. We have also analyzed their computational costs and tested the methods on spin--boson and multilevel systems.

With full bath memory and no additional total-degree truncation, the PIFFE scheme has global error $O(\Delta t^2)$ at fixed final time, assuming exact evaluation of the bath influence coefficients for the specified model. The method is deterministic and requires neither stochastic sampling nor tensor compression, so it introduces no stochastic sampling error or tensor truncation error. Numerical quadrature of the bath coefficients, when used, can be controlled separately.

PIFFE provides an intermediate approximation between QuAPI and FRODS, retaining more of the exponential influence than FRODS while expanding the far-field terms retained exponentially by QuAPI. The numerical comparisons show that its results are much closer to QuAPI than those of FRODS, even for small $r$. In particular, retaining only the self-interaction factors at $r=0$ already gives close agreement with QuAPI across the tested coupling strengths and temperatures. 

For small $r$, PIFFE requires substantially less storage than QuAPI, making much longer bath memory lengths accessible. The memory measurements and the eleven- and fifty-level examples demonstrate this advantage in practice. Truncating the bath memory and limiting the total auxiliary degree provide further reductions in both storage and computational cost. These two cutoffs offer independent controls over efficiency and accuracy, and their convergence can be checked separately from time-step refinement.

\section*{Acknowledgments}
This work was partially supported by the Academic Research Fund of the
Ministry of Education of Singapore under grant A-8002392-00-00.

\appendix
\section{Definition of the bath influence function}
\label{app:bath-influence}
Following the presentation in~\cite{sun2026}, we define $\mathcal{L}_b$ in~\eqref{eq:frodsapprox} through Wick's pairing rule. Let $\tau_k=(k-\tfrac12)\Delta t$. For nonnegative multiplicities $j_{k^-}$ and $j_{k^+}$, form the ordered list
\begin{equation}
\label{eq:appendix-discrete-bath}
  (s_1,\ldots,s_M)=\bigl(
    \underbrace{-\tau_N,\ldots,-\tau_N}_{j_{N^-}},\ldots,
    \underbrace{-\tau_1,\ldots,-\tau_1}_{j_{1^-}};
    \underbrace{\tau_1,\ldots,\tau_1}_{j_{1^+}},\ldots,
    \underbrace{\tau_N,\ldots,\tau_N}_{j_{N^+}}\bigr),
\end{equation}
where $M=\sum_{k=1}^N(j_{k^-}+j_{k^+})$.. The negative and positive times label the two branches of the density-matrix evolution.

Using the bath correlation function $B$ defined in the main text, we associate with two ordered contour times $s\leqslant s'$ the correlation
\begin{equation}
\label{eq:appendix-contour-correlation}
  C_b(s,s') =
  \begin{cases}
    B(s'-s), & s\leqslant s'\leqslant 0,\\
    B(s-s'), & 0<s\leqslant s',\\
    B(s+s'), & s\leqslant 0<s'.
  \end{cases}
\end{equation}
Since the thermal harmonic bath with linear coupling is a centered Gaussian bath, Wick's theorem gives
\begin{equation}
\label{eq:appendix-wick}
  \mathcal{L}_b(j_{N^-},\ldots,j_{1^-};j_{1^+},\ldots,j_{N^+})
  =\begin{cases}
    1, & M=0,\\
    0, & M\text{ is odd},\\
    \displaystyle\sum_{P\in\mathfrak{P}_M}
       \prod_{(a,b)\in P}C_b(s_a,s_b), & M>0\text{ is even}.
  \end{cases}
\end{equation}
Here $\mathfrak{P}_M$ is the set of all partitions of the labeled positions $\{1,\ldots,M\}$ into unordered pairs, with each pair written as $(a,b)$ with $a<b$. For positive even $M$, there are $(M-1)!!$ such pairings.

To illustrate the pairing rule, we give the following examples:
\begin{align}
  \mathcal{L}_b(1,0;1,0)
    &=C_b(-\tau_2,\tau_1)=B(-\Delta t),\\
  \mathcal{L}_b(1,1,1;1,0,1)&=0,\\
  \mathcal{L}_b(1,1;0,2)
    &=C_b(-\tau_2,-\tau_1)C_b(\tau_2,\tau_2)
      +2C_b(-\tau_1,\tau_2)C_b(-\tau_2,\tau_2)\notag\\
    &=B(\Delta t)B(0)+2B(\Delta t)B(0)
      =3B(\Delta t)B(0).
\end{align}
The last example gives $M=4$ and hence three pairings. Pairing the two occurrences at $2^+$ with each other gives the first term; pairing each of them with a negative-branch occurrence gives the other two terms. All three products equal $B(\Delta t)B(0)$.

\bibliographystyle{plain}
\bibliography{references}
\end{document}